\documentclass[fleqn,usenatbib]{mnras}

\usepackage{newtxtext,newtxmath}

\usepackage[T1]{fontenc}

\DeclareRobustCommand{\VAN}[3]{#2}
\let\VANthebibliography\thebibliography
\def\thebibliography{\DeclareRobustCommand{\VAN}[3]{##3}\VANthebibliography}

\usepackage{graphicx}	
\usepackage{amsmath}	

\usepackage[dvipsnames]{xcolor}
\usepackage{siunitx}
\usepackage{listings}

\usepackage{soul} 
\usepackage{ulem} 

\newcommand{\myvec}[1]{\boldsymbol{#1}}
\newcommand{\mymat}[1]{\mathbfss{#1}}

\DeclareSIUnit \h {\ensuremath{\mathit{h}}}
\DeclareSIUnit \parsec {pc}
\DeclareSIUnit \msol {\ensuremath{M_{\odot}}}
\DeclareSIUnit \au {AU}
\DeclareSIUnit \year {yr}

\newcommand{\rev}[1]{#1} 
\newcommand{\revcom}[1]{} 

\title[Prompt cusps \& stellar encounters]{The effect of stellar encounters on the dark matter annihilation signal from prompt cusps}

\author[J. St\"ucker et al.]{
Jens St\"ucker$^{1}$\thanks{E-mail: jstuecker@dipc.org},
Go Ogiya$^{2}$,
Simon D. M. White$^{3}$,
Raul E. Angulo$^{1,4}$
\\
$^{1}$Donostia International Physics Center (DIPC), Paseo Manuel de Lardizabal 4, 20018 Donostia-San Sebastian, Spain\\
$^{2}$Institute for Astronomy, School of Physics, Zhejiang University, Hangzhou 310027, China \\
$^{3}$Max Planck Institute for Astrophysics, Karl-Schwarzschild-Str. 1, 85741 Garching, Germany\\
$^{4}$IKERBASQUE, Basque Foundation for Science, E-48013, Bilbao, Spain
}

\date{Accepted XXX. Received YYY; in original form ZZZ}

\pubyear{2015}

\begin{document}
\label{firstpage}
\pagerange{\pageref{firstpage}--\pageref{lastpage}}
\maketitle

\begin{abstract}
Prompt cusps are the densest quasi-equilibrium dark matter objects; one forms at the instant of collapse within every isolated peak of the initial cosmological density field. They have power-law density profiles, $\rho \propto r^{-1.5}$ with central phase-space density set by the primordial velocity dispersion of the dark matter. At late times they account for $\sim 1\%$ of the dark matter mass but for $>90\%$ of its annihilation luminosity in all but the densest regions, where they are tidally disrupted. Here we demonstrate that individual stellar encounters, rather than the  mean galactic tide, are the dominant disruptors of prompt cusps within galaxies. Their cumulative effect is fully (though stochastically) characterised by an impulsive shock strength $B_* = 2\pi G\int\rho_*({\bf x}(t))\, \mathrm{d}t$ where $\rho_*$, the total mass density in stars, is integrated over a cusp's entire post-formation trajectory. Stellar encounters and mean tides have only a small effect on the halo annihilation luminosity seen by distant observers, but this is not true for the Galactic halo because of the Sun's position. For a 100~GeV WIMP, Earth-mass prompt cusps are predicted, and stellar encounters suppress their mean annihilation luminosity by a factor of two already at 20 kpc, so that their annihilation emission is predicted to appear almost uniform over the sky. The Galactic Center $\gamma$-ray Excess is thus unaffected by cusps. If it is indeed dark matter annihilation radiation, then prompt cusps in the outer Galactic halo and beyond must account for 20-80\% of the observed isotropic $\gamma$-ray background in the 1 to 10 GeV range. 
\end{abstract}




\section{Introduction}

The nature of dark matter is still unknown. One of the most popular candidates for dark matter searches is a weakly interacting massive particle (WIMP) which could be produced thermally in the early universe. If dark matter is a WIMP, it may have a significant self-interaction cross section that allows dark matter to self-annihilate in regions where the dark matter density is sufficiently high \citep{Roszkowski_2018, arcadi_2018}. 

Detection of the secondary products of such self-annihilation -- also known as indirect dark matter detection -- is one of the most promising ways of learning more about the nature of dark matter. Excitingly, the Fermi large area telescope (Fermi LAT) has detected a galactic centre excess (GCE) in gamma ray radiation in the spectral range of $1-\SI{10}{\giga \electronvolt}$ \citep{hooper_2011}. Although there has been a long-ongoing debate about the precise properties of the signal \citep[see][chapter 6, for a review]{slayter_2021}, it seems so far that the morphology and the spectrum of the signal could be consistent with a dark matter self-annihilation signal from the central regions of the Milky Way's dark matter halo \citep[e.g.][]{di_mauro_2021}. Additionally, recent high-resolution hydrodynamical simulations show that the Milky Way likely exhibits a halo with the right spatial structure to explain the GCE \citep{grand_white_2022}. However, there are also other astrophysical processes that might explain the GCE so that  it cannot yet be clearly attributed to dark matter. \rev{A few recent studies have suggested that the shape of the GCE may be better explained by templates that trace the stellar mass distribution of the Galactic bulge rather than a spherical dark matter halo} \citep{Bartels_2018, Macias_2018, Macias_2019, abazajian_2020}\rev{. However, a consensus conclusion about the morphology of the GCE remains elusive; for example,  a study by} \citet{di_mauro_2021} \rev{ assuming different background models finds consistency with radiation from a close to spherically symmetric dark matter halo.}

Recently, \citet{delos_2022_cusps} have highlighted a different aspect of the interpretation of possible indirect detection signals from the haloes of the Milky Way and of other galaxies. Recent advances in the modelling of the formation of the smallest nonlinear objects in a WIMP cosmology have led to a clearer understanding of their origin, structure and late-time abundance, leading to a re-evaluation of the expected dark matter self-annihilation signal at late times.

The primordial density field 
is smooth below the dark matter free-streaming scale  (e.g. $\sim \SI{1}{\parsec}$ comoving for a typical WIMP) and so exhibits a large number of density peaks at this scale ($\sim 10^4 - 10^5$ per solar mass of dark matter). The first nonlinear structures begin to form around redshift 30 through monolithic collapse of these peaks. Their collapse history differs radically from that of traditional, more massive haloes of the kind characterised by  \citet{nfw1996} (hereafter: NFW). While NFW haloes assemble through hierarchical accretion and merging over time-scales which are comparable to their age, a prompt cusp forms almost instantaneously at the moment of first collapse of a density peak and it contains dark particles with orbital periods much shorter than the collapse time. As a result, the density profiles of prompt cusps also differ from the NFW form, following a steep $r^{-1.5}$ density profile between an outer boundary set by the curvature of the initial density peak and an inner core determined by the physical nature of the dark matter. To distinguish these dense ``first'' objects from traditional NFW haloes, we follow  \citet{delos_2022_cusps} in referring to them as ``prompt cusps''.

\citet{delos_white_annihilation_2022} argue that prompt cusps should still be extremely abundant substructures today. In regions where their number is not significantly reduced by subsequent evolution, every solar mass of dark matter should contain tens of thousands of them, and we should expect $\sim 10^{16}$ cusps associated with the dark matter halo of the Milky Way. Since they are denser in their centres than traditional NFW haloes, prompt cusps survive tidal effects better and produce a substantially larger dark matter annihilation signal. In fact, the annihilation signal from $r^{-1.5}$-cusps is logarithmically divergent with radius, limited by the inner and outer boundaries of the power-law profile, which are set by the primordial dark matter phase-space density  \citep[e.g.][]{maccio_2012} and by the initial peak extent, respectively. This raises the signal for indirect detection compared to most previous studies. Thus \citet{delos_white_annihilation_2022} predict that these cusps should enhance the total annihilation luminosity of NFW haloes by factors ranging between $100$ and $2500$, depending on halo concentration, and additionally that they should significantly alter the morphology of the signal which, except in the densest regions, is proportional to the first power of the mean  dark matter density, rather than to its square, as usually assumed. 

As \citet{delos_white_annihilation_2022} note this ``cusp''-component impacts a possible annihilation interpretation of the GCE in two significant ways. (1)  If disruption of prompt cusps is ignored, their emission dominates that of the smooth halo component beyond about 5 degrees from the Galactic Centre, resulting in a profile in disagreement with observation. Accounting for truncation and disruption by Galactic tides reduces cusp emission from the inner Galaxy but leaves it still dominant beyond 10 degrees, reducing but not eliminating the contradiction with observation. (2) If the GCE is nevertheless due to annihilation, emission from prompt cusps in the outer Galactic halo and external to the Milky Way must constitute  a major contribution to the isotropic $\gamma$-ray background (IGRB). However, the IGRB appears to be almost completely attributable to emission from star-forming galaxies and AGN \citep{Blanco_2019} leaving little space to add an additional component. The resulting upper limit on the prompt cusp contribution to the IGRB allows \citet{delos_white_annihilation_2022} to strengthen constraints on the self-annihilation cross-section and the thermal relic mass of a hypothetical WIMP dark matter particle -- effectively excluding thermal relic WIMPs with masses $M \leq \SI{10}{\tera \electronvolt}$\footnote{This constraint is under the assumption of a bottom quark $b\overline{b}$ annihilation channel and is independent of the possible annihilation interpretation of the GCE.}.

\citet{delos_white_annihilation_2022} explicitly considered only the effect of the smooth tidal field of the Milky Way on prompt cusps.  However, stellar encounters can also be important, inducing strong impulsive shocks in dark matter substructures and potentially even disrupt them. Such encounters should be very frequent in the central region of the galaxy and therefore can affect both the predictions of the last paragraph.

The main goal of this paper is to evaluate quantitatively the effect of stellar encounters on the structure, survival and predicted annihilation signal from prompt cusps. We will find that prediction (1) is seriously affected. After accounting for the effect of stellar encounters, the tension between the GCE emission profile and that predicted including cusp emission vanishes. For prediction (2) we will find that reduced emission from cusps in the inner Galactic halo leads to slightly lower IGRB predictions for the annihilation cross-section required to produce the GCE. These predictions remain in tension with claims that the IGRB is almost entirely due to other sources. 

We note that there is already a large literature on the effect of stellar encounters on NFW subhaloes \citep{goerdt_2007, angus_2007, green_2007, schneider_2010, Delos_2019, kavanagh_2021, shen_2022, facchinetti_2022}. \revcom{Added Facchinetti 2022}  Much of this is based on the incorrect assumption that systems disrupt when the total injected energy exceeds their initial binding energy, and hence needs to be read with care \citep{aguilar_1985,vandenbosch_2018}. A particularly clear and general treatment has been presented by \citet{Delos_2019} and we will often refer to this article as representative of stellar encounters with NFW haloes. Such results cannot be applied to prompt cusps, since their power-law structure dramatically enhances their resilience to tidal effects in comparison  to the inner regions of NFW profiles \citep[see][ for a discussion of different power-law profiles]{stuecker_2022}. The only previous study to consider the effect of stellar encounters on prompt cusps is \citet{Ishiyama_2010}, but unfortunately this presented a very limited treatment and used incorrect assumptions when scaling with encounter strength, leading to the erroneous conclusion  that cusps would never be disrupted by stellar encounters. We will discuss this in more detail below.

The structure of this paper is as follows. In Section \ref{sec:theory} we briefly introduce both the theoretical basis for predicting the distribution of prompt cusp structural properties and the physical formalism needed to describe the effects of their encounters with stars. We also present a novel, simple and very general scheme for calculating the full distribution of impulsive stellar shocks experienced by cusps (or normal subhaloes) as they pass through a  galaxy (e.g. through the Milky Way's disk or bulge). In Section \ref{sec:milkyway} we numerically integrate cusp orbits in a realistic Milky Way model, and we infer the parameters describing their shock histories. In Section \ref{sec:simulations} we use idealized N-body simulations to estimate the disruptive effects of stellar encounters, developing simple formulae that predict the structure and annihilation signal of cusps that have gone through arbitrarily many encounters. Additionally, we show how this can be supplemented to include the effects of stripping by the smooth Galactic tidal field. In Section \ref{sec:results} we present our main results, predictions for the profile of the prompt cusp contribution to the annihilation signal of the Galactic halo both as seen by a distant observer and as seen from the Earth, together with an assessment of how prompt cusps affect the form and relative amplitude of the GCE and the IGRB. In Section \ref{sec:conclusion} we will discuss the implications of these findings for indirect dark matter searches.

We make almost all codes used in this study available through an online \textsc{python} repository\footnote{\href{https://github.com/jstuecker/cusp-encounters}{https://github.com/jstuecker/cusp-encounters}} so that our methods can easily be used in future studies and the results of this paper can be reproduced independently.

\section{Theory} \label{sec:theory}

As described in the introduction, the effect of stellar encounters on NFW subhaloes has already been studied extensively, although in many cases using an incorrect criterion for subhalo disruption. There has, however, been no realistic study of the effects for power-law cusps. Furthermore, even in the  NFW case, most studies have not realistically modelled the full distribution of encounter parameters. Here we present the theoretical considerations necessary to treat full encounter histories in an accurate, general but simple way.

For this we present in Section \ref{sec:cusps} the distribution of initial cusp properties as derived from the statistics of peaks in the initial gaussian density field, in Section \ref{sec:phasespacecores} how cusp structure can be modified by an inner core to account for the upper limit on phase-space density, in Section \ref{sec:impulsiveencounters} the impulsive shocks induced by stellar encounters, in Section \ref{sec:encounternumber} a calculation of the total number of encounters expected on passing through a star distribution with arbitrary stellar mass and velocity distributions, in Section \ref{sec:shockhistories} the statistical distribution of shock histories that follows from these considerations.

\subsection{Cusps} \label{sec:cusps}
Several numerical studies have found that peaks on the dark matter free-streaming scale collapse promptly to form dense cusps \citep{diemand_2005, Ishiyama_2010, Anderhalden_2013, Ishiyama_2014, Polisensky2015, angulo_2017, Ogiya2018, Delos_2018_ultracompact, delos_2018_ultracompact_constraints, delos_2019_predicting, colombi_2021, delos_2022_cusps, White_2022}. These cusps have a density profile,
\begin{align}
    \rho(r) &= A r^{-3/2}, \label{eqn:cuspdensity}
\end{align}
parameterised by a normalization $A$ and an outer radius $r_{\rm{cusp}}$ which limits the extent of the profile. \citet{delos_2022_cusps} find that both parameters can be predicted well for a given cusp from the properties of the initial density peak from which it forms. Specifically, 
\begin{align}
    A &= 24 \rho_0 a_{\rm{col}}^{-1.5} R^{1.5}, \label{eqn:Acusp} \\
    r_{\rm{cusp}} &= 0.11 a_{\rm{col}} R ,\label{eqn:rcusp}
\end{align}
where $\rho_0$ is the mean dark matter density of the universe today, $a_{\rm{col}}$ is the scalefactor when the peak first collapses and $R$ is the Lagrangian size of the initial density peak defined as $R = \sqrt{|\delta / \nabla^2 \delta|}$, where $\delta({\bf x})$ is the linear overdensity field as a function of comoving position. The time of collapse can be estimated to sufficient accuracy by an ellipsoidal collapse model based on the triaxial structure of the initial peak. A more detailed description can be found in \citet{delos_white_annihilation_2022}.

We follow the descriptions of \citet{delos_white_annihilation_2022} to sample a distribution of cusps. For this we use a dark matter power spectrum generated by the Boltzmann code \textsc{CLASS} \citep{class_paper_2011} up to a resolved wavenumber of $k = \SI{e4}{\h \per \mega \parsec}$ at $z=30.6$. Beyond that scale we use the analytic prescriptions of \citet{Hu_1996}, but normalized so that it matches the \textsc{CLASS} spectrum at $k = \SI{e4}{\h \per \mega \parsec}$. We multiply this spectrum using the exponential power spectrum cutoff description of \citet{bertschinger_2006} for a WIMP with mass $m_{\rm{WIMP}} = \SI{100}{\giga\electronvolt}$ and decoupling temperature $T_{\rm{d}} = \SI{30}{\mega \electronvolt}$ (corresponding to a decoupling scale-factor of $a_{\rm{d}} = \SI{5.33e-12}{}$). We use the free streaming length $k_{\rm{FS}} = \SI{1.06}{\per \parsec}$ as calculated by \citet{delos_white_annihilation_2022}. The distribution of initial density peaks can be sampled analytically given the initial power spectrum as described by \citet{bbks_1996}. We have created our own implementation of the sampling of peaks which is published in the code repository of this paper. However, we tested it against the peak sampling implementation that was published by \citet{delos_2019_predicting}. We map the distribution of peaks onto a distribution of cusps by using equations \eqref{eqn:Acusp} -- \eqref{eqn:rcusp} with the ellipsoidal collapse correction for $a_{\rm{col}}$ as explained by \citet{delos_white_annihilation_2022}  based on the approximation from \citet{sheth_2001}.

\begin{figure}
    \centering
    \includegraphics[width=\columnwidth]{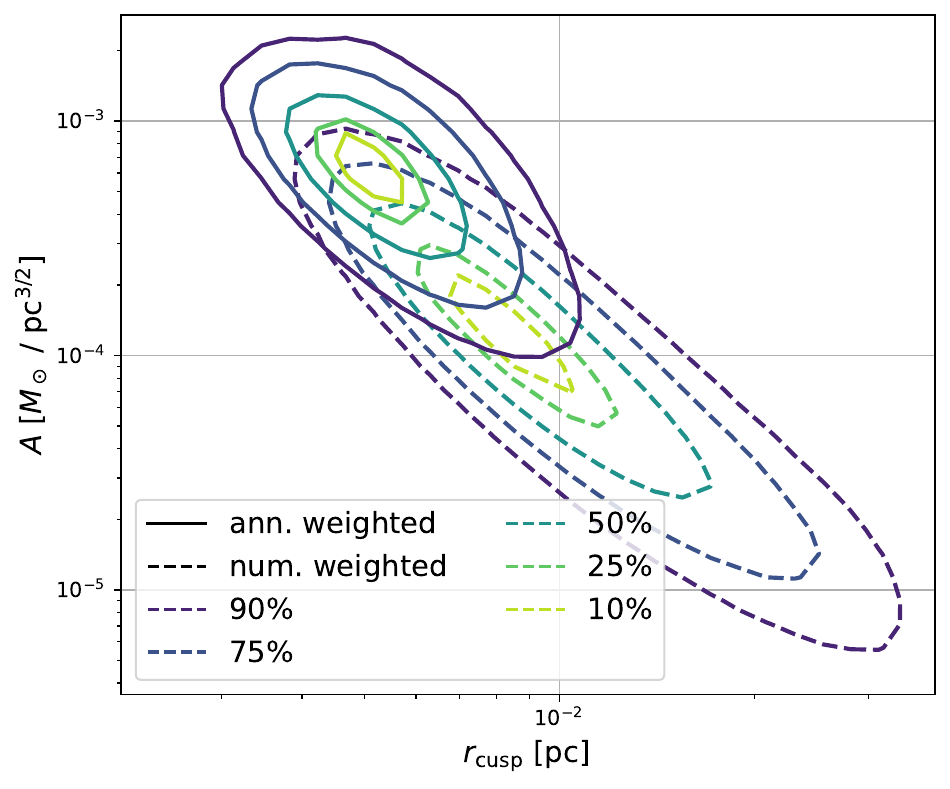}
    \caption{The distribution of the cusp normalization $A$ versus the cusp truncation radius $r_{\rm{cusp}}$. Typical cusps that dominate the annihilation rate distribution have $A \sim \SI{8e-4}{\msol \parsec^{-3/2}}$ and $r_{\rm{cusp}} \sim \SI{5e-3}{\parsec}$.}
    \label{fig:cusp_A_distribution}
\end{figure}

We show the resulting distribution of cusps in Figure \ref{fig:cusp_A_distribution} (dashed contours). However, more relevant than the number distribution of cusps is the annihilation weighted distribution. Under the assumption of a velocity independent cross section, the annihilation rate of a cusp should be proportional to
\begin{align}
    J &= \int_{r_{\rm{core}}}^{r_{\rm{cusp}}} \rho^2 \, \mathrm{d}^3 r \nonumber \\
      &= 4 \pi A^2 \log(r_{\rm{cusp}} / r_{\rm{core}}),
      \label{eq:j_powerlaw}
\end{align}
where we have neglected any contributions from $r < r_{\rm{core}}$ and $r > r_{\rm{cusp}}$ where $r_{\rm{core}}$ is the core radius enforced by the phase-space density constraint \citep{delos_2022_cusps}. We will explain in Section \ref{sec:phasespacecores} how to calculate and treat the core radius more precisely.

The weighted distributions are shown as solid contours in Figure~\ref{fig:cusp_A_distribution}. We can see that for this specific dark matter particle, the typical cusp relevant for the annihilation rate has  $A \sim \SI{8e-4}{\msol \parsec^{-3/2}}$ and $r_{\rm{cusp}} \sim \SI{5e-3}{\parsec} \approx \SI{e3}{\au}$ (in physical units). It collapses at redshift $z\sim 30.$

\subsection{Phase-space cores} \label{sec:phasespacecores}
The fine-grained phase-space density of dark matter is constant as a function of time (Liouville's theorem). The coarse-grained phase-space density -- defined as the average over some finite phase-space volume -- can therefore never exceed the fine-grained value established at dark matter freeze-out \citep{tremaine_gunn}. Let us denote the maximum of this fine-grained phase-space density by $f_{\rm{max}}$. The value of $f_{\rm{max}}$ is set in the early universe and depends strongly on the type of dark matter considered.  For the WIMP model considered above, for example, it is $f_{\rm{max}} = 9.98 \cdot 10^{13} \frac{\SI{}{\msol}}{\SI{}{\kilo \metre^3 \second^{-3}\parsec^3}}$, whereas for thermal relic warm dark matter with $m_X = 3.5 \rm{keV}$ it is $f_{\rm{max}} = 6.75 \cdot 10^{-2} \frac{\SI{}{\msol}}{\SI{}{\kilo \metre^3 \second^{-3}\parsec^3}}$ \citep[see e.g.][]{delos_2022_cusps}\footnote{Our values here differ slightly from the values of \citep{delos_2022_cusps} since we use $\Omega_x = 0.26$ instead of $0.3$ for the dark matter density parameter.}.

The power-law profile of equation \eqref{eqn:cuspdensity} corresponds, for an isotropic velocity distribution, to a phase-space distribution function,
\begin{align}
    f(E) &= f_0 E^{-9/2}, \label{eqn:plawphasespace} \\
    f_0 &= \frac{1120 \sqrt{2} \pi^{2} A^{4} G^{3}}{9}, \label{eqn:f0}
\end{align}
\citep[see e.g.][]{stuecker_2022}. This description must break down at energies where $f(E) > f_{\rm{max}}$. To estimate the radial scale where this happens we can consider at each radius the largest reachable phase-space density -- given by $f(\phi(r))$ where $\phi(r)$ is the potential. Inserting this into equation \eqref{eqn:plawphasespace} and inverting for r we find
\begin{align}
    r_{\rm{core}} &= \frac{3 \sqrt[9]{3} \cdot 70^{\frac{4}{9}}}{64 \pi^{\frac{10}{9}} A^{\frac{2}{9}} G^{\frac{2}{3}} f_{\rm{max}}^{\frac{4}{9}}} \nonumber \\
                  &\approx \SI{1.03e-5}{\parsec} 
                  \left(\frac{A}{\SI{e-3}{\msol \parsec^{-\frac{3}{2}}}}\right)^{-\frac{2}{9}}
                  \left(\frac{f_{\rm{max}}}{10^{14}  \frac{\SI{}{\msol}}{\SI{}{\kilo \metre^3 \second^{-3}\parsec^3}} } \right)^{-\frac{4}{9}}.
\end{align}
Thus in the above WIMP model, typical cusps relevant for the annihilation calculation  have thermal core size, $r_{\rm{core}} \sim \SI{e-5}{\parsec} \sim \SI{2}{\au}$.

The detailed shape of the density profile near the core radius is unclear. It would be desirable to run simulations with the actual primordial velocity distribution of a WIMP, so that phase-space cores are created self-consistently, and then to measure their density profile. Such simulations would be  computationally demanding and have not yet been performed, but they are not far beyond current capabilities. (Consider that typically $r_{\rm{cusp}} \sim 500 r_{\rm{core}}$ so that simulations that resolve the cusp profile well are typically only an order of magnitude in linear scale from the core radius.) 

To obtain a profile that has the desired maximum phase-space density and connects smoothly to the appropriate  power-law at larger radii we make the following Ansatz for the phase-space density,
\begin{align}
    f(E) &= \frac{f_0}{(E + E_{\rm{core}})^{9/2}}, \label{eqn:phasespacecored}
\end{align}
where $E_{\rm{core}}$ is the core energy scale where the phase-space density $f_{\rm{max}}$ would be reached in the fiducial power-law profile:
\begin{align}
    E_{\rm{core}} &= \left( \frac{f_0}{f_{\rm{max}}} \right)^{2/9},
\end{align}
so that together with equation \eqref{eqn:f0} the profile is fully specified through $A$ and $f_{{\rm{max}}}$. Note that for $f_{\rm{max}} \rightarrow \infty$ the pure power-law profile is recovered.

\begin{figure}
    \centering
    \includegraphics[width=\columnwidth]{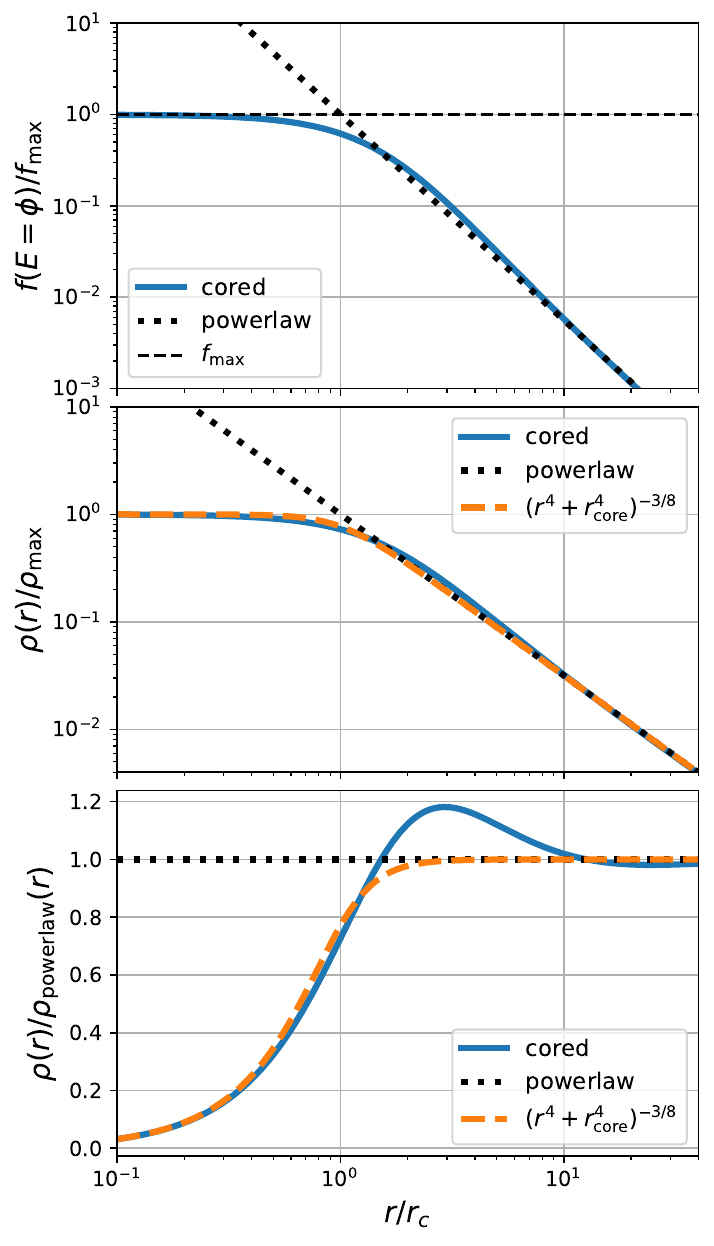}
    \caption{The phase-space and density profiles of a $r^{-3/2}$ cusp with a core induced by the phase-space constraint $f < f_{\rm{max}}$. The top panel shows the phase-space density $f(\phi(r))$ corresponding to the highest phase-space density present at each radius. The central panel shows the density profile and the bottom panel shows the density profile divided by the fiducial power-law profile. In each case, we see a rapid transition between the power-law behaviour and a maximum central value.}
    \label{fig:cored_profile}
\end{figure}

We can integrate over the velocity components of the phase-space density to find the density,
\begin{align}
    \rho(r) &= \int f(E) \,\mathrm{d}^3 v \nonumber \\
            &= 4\pi \int_\phi^\infty \int_0^{r\sqrt{2(E - \phi)}}  \frac{L f(E)}{r^2 \sqrt{2E - 2\phi - \frac{L^2}{r^2}}} \, \mathrm{d}L \, \mathrm{d}E  \nonumber \\
            &= 4\pi  \int_\phi^\infty  f(E) \sqrt{2(E - \phi)} \, \mathrm{d}E  \nonumber  \\
            &= \frac{64 \pi \sqrt{2} }{105 \left(E_{\rm{core}} + \phi(r)\right)^{3}},
\end{align}
as a function of the potential, $\phi$, which is normalized so that $\phi(0) = 0$ and $\phi(r \rightarrow \infty) \rightarrow \infty$. Combined with Poisson's equation this forms a nonlinear second order differential equation for the potential,
\begin{align}
    \frac{\partial_r (r^2 \partial_r \phi)}{4 \pi G r^2} &= \frac{64 \pi \sqrt{2} }{105 \left(E_{\rm{core}} + \phi(r)\right)^{3}}.
\end{align}
We do not know how to solve this differential equation analytically and therefore solve it through numerical integration starting at $r=0$ using $\phi(0) = \partial_r \phi(0) = 0$. As a result we find $\rho(r)$ and $\phi(r)$ and we display these and the phase-space density $f(\phi(r))$ in Figure \ref{fig:cored_profile}. 

As expected, the density profile reaches a well defined maximum at $r \lesssim r_{\rm{core}}$, given by 
\begin{align}
    \rho_{\rm{max}} &= A r_{\rm{core}}^{-3/2} \nonumber \\
                    &\propto A^{4/3} f_{\rm{max}}^{2/3}.
\end{align}

A good approximation to the density profile is given by
\begin{align}
    \rho(r) &= A (r^4 + r_{\rm{core}}^4)^{-3/8},
\end{align}
which we also show as an orange line in Figure \ref{fig:cored_profile}. This deviates at most by $20\%$ from the actual density around $r \sim 3r_c$ where the latter is slightly enhanced with respect to the fiducial power-law. This enhancement is similar to the well-known behavior of the isothermal sphere where the cored solution rises above the asymptotic $r^{-2}$ power-law solution at radii comparable to the core radius before aymptoting to constant density at smaller radii. \citep[e.g.][]{BinneyTremaine2008}.

We note that while there are no simulations which have a phase-space density constraint consistent with the expected free-streaming scale, there have been simulations by \citet{maccio_2012} with artificially large initial velocity dispersion and hence a lowered upper limit on phase-space density. Although it is difficult to compare these directly with our profile, it seems that the final cores do saturate the phase-space bound and that they transition relatively quickly to the asymptotic power-law behaviour -- both consistent with the profile that we propose here.

We find that the annihilation radiation from the cored profile inside radius $r$ can be approximated for $r > 10 r_{\rm{core}}$ by
\begin{align}
    J(<r) &= 4 \pi A^2 (0.531 + \log(r / r_{\rm{core}})). \label{eqn:Jinitial}
\end{align}
This is marginally larger than the annihilation radiation one obtains when assuming the power-law profile down to $r_{\rm{core}}$ (compare equation \eqref{eq:j_powerlaw}) and a uniform density at smaller radii, in which case the $0.531$ gets replaced by $0.333$. This is due to the slight enhancement of the profile around $3 r_{\rm{core}}$.

When needed, we will use the numerical cored profile throughout this study.

\subsection{Impulsive encounters} \label{sec:impulsiveencounters}
We consider a star with mass $M_*$ passing a cusp on a linear orbit at constant relative velocity $v$ and minimal distance $b$. We can approximate the tidal forces acting on cusp particles through a multipole expansion of the potential up to second order,
\begin{align}
    \phi(\myvec{x}, t) &= \phi_s(\myvec{x}-\myvec{x}_0,t ) - \mymat{T}_0(t) (\myvec{x} - \myvec{x}_0),
\end{align}
where $\phi_s$ denotes the self-potential of the cusp, $\myvec{x}_0$ is the location of centre of the cusp and $\mymat{T}_0(t)$ is the tidal field of the star evaluated at $\myvec{x}_0$. Here we have neglected zeroth- and first-order terms, since they do not affect the internal dynamics of the cusp \citep[e.g.][]{stuecker_2021_bp}. This ``distant-tide'' approximation is valid for particles that are close enough to the centre, $||\myvec{x} - \myvec{x}_0||  \ll b$ . We will see that for encounters with stars with masses of order $\SI{}{\msol}$ the distant tide approximation is excellent for all particles that remain bound to the cusp (which typically have small $\Delta \myvec{x}$) and is only  violated for particles that are kicked so strongly that they will leave the system. Thus, it is safe to adopt this approximation in all our calculations.

Further, we can assume the impulsive approximation which considers the limit that particles move very little within the cusp during the time of the encounter. In this case, the total change in the velocity of a particle due to the encounter can be approximated by
\begin{align}
    \Delta v &= \left(\int \mymat{T}_0(t) \mathrm{d}t \right)  (\myvec{x} - \myvec{x}_0 ) \nonumber \\
             &= \mymat{K} (\myvec{x} - \myvec{x}_0), \label{eqn:kick}
\end{align}
where we have defined the shock tensor $K$. 

It is easy to see that the impulsive approximation is an excellent approximation here. The dynamical time-scale at radius $r$ of our cusp is given by 
\begin{align}
    t_{\rm{d}}(r) &= \frac{r}{v_{\rm{circ}} (r)} \nonumber \\
           &= \sqrt{\frac{r^3}{G M(<r)}},
\end{align}
where $M(<r)$ is the enclosed mass profile. The internal dynamical time-scale is shortest at the core-radius, where it is of order $\SI{2e4}{\year}$ for the strongest cusps (which dominate the annihilation distribution). The impact parameter of the weakest relevant encounters is of order $b \sim \SI{e4}{\au}$ with $v \sim \SI{200}{\kilo \metre \per \second}$ which gives an encounter time-scale of order $t_{\rm{enc}} = b / v \sim \SI{200}{\year}$ which is two orders of magnitude smaller than the dynamical time scale of the quickest particles in the cusp. Stronger  encounters (which are more relevant) will happen on even shorter time-scales and most particles orbit on longer time scales so that the ratio should be even larger in practice and we can safely assume the impulsive limit for all of our calculations.

If we assume that the star is a point mass with tidal field
\begin{align}
    T_{ij}(t) &= - \partial_i \partial_j \left( - \frac{M_* G}{\lVert \myvec{x} - {x_*}(t) \rVert } \right),
\end{align}
and we assume its trajectory (without loss of generality) to be along the y-direction with the closest encounter at the coordinate $(b, 0, 0)$, then the shock tensor is given by
\begin{align}
    \mymat{K} &= B \begin{pmatrix} 1 & 0 & 0\\
                           0 & 0 & 0\\
                           0 & 0 & -1 \end{pmatrix} \label{eqn:kicktensor} \\
    B &= \frac{2 G M_*}{v b^2} \label{eqn:shockparameter}
\end{align}
\citep[e.g.][]{aguilar_1985} where we have defined the shock parameter $B$.  We note that $B$ has dimensions of the inverse of time, but to simplify intuitive understanding we will typically state it in units of $ \SI{}{\kilo \metre \per \second \per \parsec}$.

It is clear that the only relevant parameter for describing the effect of a distant and impulsive  stellar tidal shock on a prompt cusp is the tidal shock parameter $B$. The individual values of $b$, $M_*$ and $v$ matter only insofar that they determine the value of $B$. 

\begin{figure}
    \centering
    \includegraphics[width=\columnwidth]{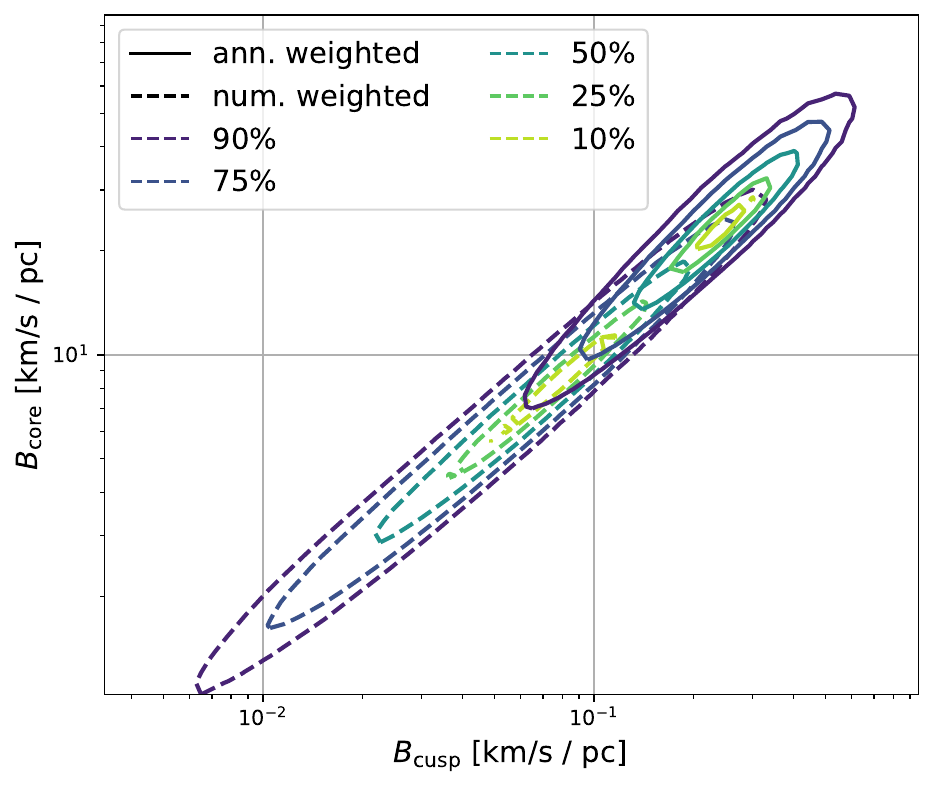}
    \caption{The distributions of $B_{\rm{core}}$ and $B_{\rm{cusp}}$ which indicate the resilience of prompt cusps against tidal shocks. A tidal shock with $B \gtrsim B_{\rm{cusp}}$ will likely affect the cusp's profile significantly. Shocks with $B \gtrsim B_{\rm{core}}$ will lead to disruption.}
    \label{fig:cusp_B_distribution}
\end{figure}

We can get a feeling for what range of $B$ values will be relevant for typical cusps by considering the values, 
\begin{align}
    B_{\rm{cusp}} &= \frac{v_{\rm{circ}}(r_{\rm{cusp}})}{r_{\rm{cusp}}} = \sqrt{\frac{8 \pi G A}{3 r_{\rm{cusp}}^{3/2}}}, \\
    B_{\rm{core}} &= \frac{v_{\rm{circ}}(r_{\rm{core}})}{r_{\rm{core}}} = \sqrt{\frac{8 \pi G A}{3 r_{\rm{core}}^{3/2}}} \propto  A^{2/3} f_{\rm{max}}^{1/3}.
    \label{eqn:Bcusp_core}
\end{align}
$B_{\rm{cusp}}$ indicates the strength of a tidal shock that is needed to induce a velocity change at the radius $r_{\rm{cusp}}$ as large as the circular velocity and $B_{\rm{core}}$ is the analogue quantity at the core radius. We show the distributions of these two parameters for the fiducial $\SI{100}{\giga\electronvolt}$ WIMP in Figure \ref{fig:cusp_B_distribution}. We can expect that tidal shocks with $B \gtrsim B_{\rm{cusp}}$ will significantly alter the profile of the cusp. Tidal shocks with $B \gtrsim B_{\rm{core}}$ may possibly lead to complete disruption. Shocks with $B \ll B_{\rm{cusp}}$ will leave the cusp largely unaffected. Typical cusps that are relevant for the annihilation rate have values of $B_{\rm{cusp}}$ and $B_{\rm{core}}$ of order $\SI{0.3}{\kilo \metre \per \second \per \parsec}$ and $\SI{30}{\kilo \metre \per \second \per \parsec}$ respectively. The impact parameters that are needed to reach such shock parameters for $M_* = \SI{1}{\msol}$ and $v =  \SI{200}{\kilo \metre \per \second}$ are $\SI{1e-2}{\parsec} \approx \SI{2000}{\au}$ and $\SI{1e-3}{\parsec} \approx \SI{200}{\au}$ respectively. We will see that tidal shocks of this order are not only possible, but also quite likely. It is therefore important to make precise quantitative calculations to evaluate the number of such encounters and the effect they have on the aggregate annihilation luminosity from cusps.

Finally, it is useful to introduce the characteristic spatial scale associated with the action of shocks on a cusp,
\begin{align}
    r_B &= \left(\frac{8 \pi G A}{3 B^2 }\right)^{2/3} \nonumber \\
        &= \left( \frac{2 \pi A}{3 G} \right)^{2/3} \frac{b^{8/3} v^{4/3}}{M_* ^{4/3}},
\end{align}
which is the radius where the change in velocity induced by the shock is of order the circular velocity. We will see in Section \ref{sec:simulations} that the majority of particles beyond $r_B$ will leave the system. The distant tide approximation is a good approximation if $\text{min} ( r_B , r_{\rm{cusp}}) \ll b$ so that it holds for all particles that remain bound. We find that typical encounter scenarios with stars for prompt cusps have $\text{min} (r_B , r_{\rm{cusp}}) \ll 0.1 b$ for any impact parameter $b$, so that the distant tide approximation is always valid. While we do not focus on NFW haloes in this study, we note that many of our subsequent derivations are of interest for studies of NFW substructures as well. For these, we estimate that $r_{B, \rm{NFW}} < 0.1 b$ (defined through the circular velocity criterion for an NFW profile) holds for typical closest impact parameters with $b \lesssim \SI{1000}{AU}$ for haloes with virial masses $m_{200\rm{c}} \lesssim 1 M_\odot$ (and concentration $c \sim 30$). Many of our results below can therefore also be applied to NFW subhaloes with masses below $1.0 M_\odot$, but additional care must be taken if larger masses are considered since the distant tide approximation may then fail.

The goal of the remaining parts of this section is to determine the distribution of shock parameters $B$ for a cusp that is moving through an arbitrary stellar distribution (e.g. a component of the Milky Way). We note that this calculation is both simpler and more accurate than previous calculations in the literature which have focused on estimating the distribution of the impact parameter $b$. 

\subsection{The expected number of encounters} \label{sec:encounternumber}
We want to estimate the expected number of encounters with a tidal shock parameter that is greater than $B$, $N(>B)$, for a cusp that is orbiting through an arbitrary stellar distribution. The stellar distribution can be described by a mass dependent phase-space (number) density
\begin{align}
    f_*(\myvec{x}, \myvec{v}, M) = \frac{\mathrm{d}N}{\mathrm{d}^3 x\,\mathrm{d}^3 v\, \mathrm{d} M},
\end{align}
which is normalized so that an integral over a phase-space volume and over a stellar mass interval gives the number of stars within that volume and mass range. An integral over phase space alone gives the stellar mass function, which is allowed to vary spatially. 

If our cusp was passing through a uniform medium without velocity dispersion and with number density $n_*$ then the expected number of encounters in a time-interval $\mathrm{d}t$ with impact parameters in the range $[b, b + \mathrm{d}b]$ is given by
\begin{align}
    \mathrm{d} N &=  2 \pi n_*  b v \, \mathrm{d}b \, \mathrm{d}t \label{eqn:uniformenc},
\end{align}
where $v$ is the relative velocity with respect to the medium. Equation \eqref{eqn:uniformenc} arises by considering stars when they get closest to our cusp, which is when they pass through the plane orthogonal to the velocity vector. The relevant velocity here is the relative velocity between the stars and the cusp; higher velocities lead to more frequent encounters.

For arbitrary phase-space distributions we have to consider each velocity and mass bin individually. The number of encounters within time interval $\mathrm{d} t$, impact parameter range $(b, b+\mathrm{d} b)$ and  encounter velocity bin $\mathrm{d}^3 v$ with stars that have masses in the range $(M, M + \, \mathrm{d} M)$ is given by
\begin{align}
    \mathrm{d} N &= (2 \pi b \, \mathrm{d} b) (f_*(\myvec{x}, \myvec{v} - \myvec{v_{\mathrm{rel}}}, M) \, \mathrm{d}^3 v \, \mathrm{d} M) (v \, \mathrm{d} t), \label{eqn:differentialencounters}
\end{align}
where $\myvec{v}$ is the encounter velocity, and $v_{\rm{rel}}$ is the relative velocity between our cusp and the zero-point of the stellar phase-space distribution. Here, we have assumed that the phase space distribution function does not vary significantly over the distance $b$. This is an excellent approximation, since typical encounters of interest will have $b \ll \SI{1}{\parsec}$ whereas the stellar distribution varies on much larger scales ($\gg \SI{100}{\parsec}$).

Now, it would be straightforward to estimate, for example, the total number of encounters with impact parameters smaller than some given $b$ by integrating \eqref{eqn:differentialencounters} over the corresponding variables. In general this would turn out to depend both on the phase-space distribution of the stars and the stellar mass function.

However, as discussed in the previous subsection, we are actually interested in the distribution of shock parameters $B$. This can be evaluated by integrating \eqref{eqn:differentialencounters} under the constraint that $B = GM/vb^2$ which leads us to
\begin{align}
   \frac{\mathrm{d}N}{\mathrm{d}B} &=  \int_{6 \rm{dim.}} \delta_D \left(\frac{2MG}{v b^2}  - B\right) \frac{\mathrm{d}^6 N}{\mathrm{d}^3v \, \mathrm{d}b \, \mathrm{d}M \, \mathrm{d}t}   \mathrm{d}^3 v \, \mathrm{d} b \, \mathrm{d} M \, \mathrm{d} t,
\end{align}
where $\delta_D$ is the Dirac delta function. We evaluate the integral, integrating in $b$ first:
\begin{align}
    \frac{\mathrm{d}N}{\mathrm{d}B} &=  \int  \int  \int f_*(\myvec{x}, \myvec{v} - \myvec{v_{\rm{rel}}}, M) v g(M,B,v) \, \mathrm{d}^3v  \, \mathrm{d}t \, \mathrm{d}M, \\
    g(M,B,v) &= \int_0^\infty 2\pi b \delta_D \left(\frac{2MG}{v b^2}  - B\right) \mathrm{d}b \nonumber \\
             &= \int_0^\infty 2\pi \frac{MG}{v B'^2} \delta_D \left(B'  - B\right)  \mathrm{d}B' \nonumber\\
             &= \frac{2 \pi M G}{v B^2},
\end{align}
where we have used the substitution $B'  = \frac{2MG}{v b^2}$ to evaluate the integral.

Sorting and evaluating individual terms gives us
\begin{align}
    \frac{\mathrm{d}N}{\mathrm{d}B} &=  \frac{2 \pi G}{B^2} \int \int \int  M f_*(\myvec{x}, \myvec{v} - \myvec{v_{\rm{rel}}}, M) \label{eqn:dndbintegrals} \, \mathrm{d}^3v \, \mathrm{d}M \,\mathrm{d}t \nonumber \\
                  &= \frac{2 \pi G}{B^2} \int \rho_*(\myvec{x}(t)) \, \mathrm{d}t \nonumber \\
                  &= \frac{2 \pi G}{B^2} \chi_*,
\end{align}
where we have used the fact that the mass weighted integral over the phase-space number density gives the stellar mass density $\rho_*$. Further we have defined the time integral of the stellar density along the cusp trajectory $\chi_*$. It is further convenient to define the characteristic shock parameter,
\begin{align}
    B_* &= 2 \pi G \chi_*,
\end{align}
so that
\begin{align}
    \frac{\mathrm{d}N}{\mathrm{d}B} &= \frac{B_*}{B^2}, \\
    N(>B) &= \frac{B_*}{B}.
\end{align}
Therefore, we expect on average one encounter with $B > B_*$.

It is worth noting that this result is surprisingly independent of the phase-space distribution function of stars and the stellar mass function, but depends only on the stellar mass density. The mass function is irrelevant, because at a fixed mass density, a reduction in mass leads to an increase in number density, thus making close encounters more likely to the same degree that it decreases the strength of such encounters.  A similar coincidence holds for the velocity dependence. The number of stars that are encountered increases with the velocity, while at the same time the weight of each encounter decreases with the velocity to such a degree that the two effects cancel exactly. We call these effects the {\it encounter conspiracy}.

It is clear that these two simplifications could, in principle,  break down at some scale. Mass function independence breaks down if the distant tide approximation fails -- if we make our perturbers less massive, the encounters get closer for a given value of $B$. For very small masses e.g. $\SI{e-6}{\msol}$ the closest approach needed for a significant perturbation would be of order one $\SI{}{AU}$, smaller than core radius of a cusp; the distant tide approximation would then certainly fail. The integral over the mass function in equation \eqref{eqn:dndbintegrals} should thus have a lower limit, in principle. The independence of velocity breaks down for very small encounter velocities. If an encounter takes longer than the orbital times within the cusp, then the cusp will react adiabatically, with no long-term changes in energy except for particles that leave the system in the adiabatic limit. Thus our approximations fail for stars that are moving almost at the same velocity as the cusp. However, neither of these problems has a significant effect in practice, since almost all stellar mass is in objects of mass within an order of magnitude or so of $1.0~\SI{}{\msol}$ and very few encounter velocities are smaller than, say,  $\SI{10}{\kilo \metre \per \second}$. 

We note that the effects of encounters with other massive objects, such as planets or other prompt cusps would be overestimated if the calculation of this section were applied. However, the mass density in planets is so much lower than that in stars that planets are quite irrelevant in this context. The mass density in prompt cusps is non-negligible at large halocentric radii, but when their extended profile is taken into account, we find that the strongest possible shocks are far below $B \ll \SI{e-4}{\kilo\metre \per\second \per \parsec}$ so that cusps cannot significantly shock other cusps (cf. Figure \ref{fig:cusp_B_distribution}). Therefore, stars pose the only significant contributor to the distribution of encounter shocks.

\subsection{Shock Histories} \label{sec:shockhistories}
We assume that all aspects of the problem follow Poisson statistics -- for example that stars are drawn through a Poisson process from the continuous phase-space distribution and that stellar masses are drawn through a Poisson process from the stellar mass function. Then also the shock parameter distribution has to follow Poisson statistics. That means the probability of having exactly $k$ encounters with shock strength bigger than $B$ is given by
\begin{align}
    F(k, B) &= \frac{\left(B_* / B \right)^k \exp(- B_* / B) }{k!}.
\end{align}
In particular, the probability of having at least one encounter with shock strength bigger than $B$ is given by
\begin{align}
    F(\geq 1, B) &= 1 - F(0, B) \nonumber \\
                 &= 1 - \exp(- B_* / B).
\end{align}
The probability density function of the strongest encounter is therefore
\begin{align}
    f_1(B) &= \frac{\mathrm{d} F(\geq 1, B)}{\mathrm{d} B} \nonumber \\
           &= \frac{B_*}{B^2} \exp(- B_* / B). \label{eqn:strongestencounter}
\end{align}
It is straightforward to derive the corresponding functions for the 2nd, 3rd etc strongest shock. However, the distribution of e.g. the strongest and the second strongest shock are not independent and parameterising the joint distribution is rather cumbersome. When considering a large number of encounters it is more convenient to draw actual realisations. This can easily be done by mapping $B$ onto another random variable that follows a uniform distribution $x := B_* / B$,
\begin{align}
    \frac{\mathrm{d} N}{\mathrm{d} x} &= \frac{\mathrm{d} N}{\mathrm{d} B} \frac{\mathrm{d} B}{\mathrm{d} x}  = 1.
\end{align}

\begin{figure}
    \centering
    \includegraphics[width=\columnwidth]{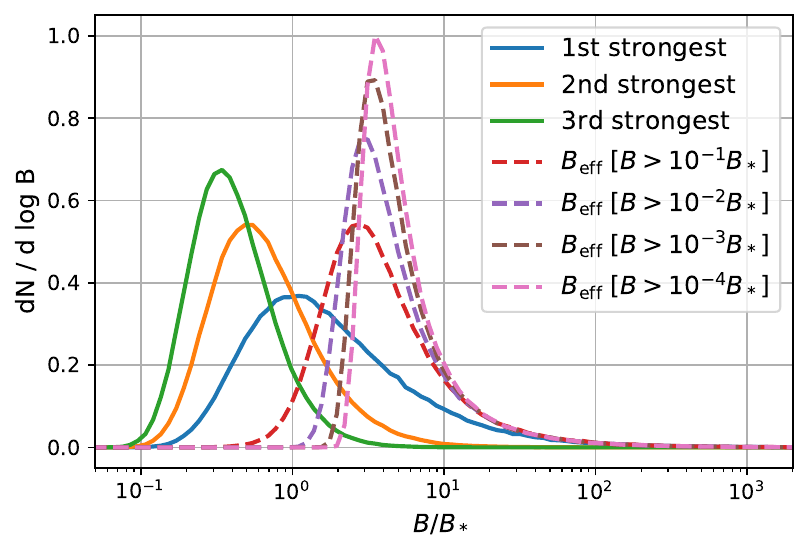}
    \caption{The distribution of shock parameters. The blue, orange and green lines show the distribution of the strongest, 2nd strongest and 3rd strongest shock. The dashed lines show the distribution of the effective shock parameter according to equation \eqref{eqn:Beff} when truncating the (divergent) distribution at different values of $B_{\rm{min}}$.}
    \label{fig:shockdistribution}
\end{figure}
We can then create a realization of a history of shocks with $B > B_{\rm{min}}$, by sampling $k$ uniformly distributed random variables $x_i$ on the interval $[0, B_* / B_{\rm{min}}]$ and then transforming them as $B = B_* x^{-1}$. The number $k$ must itself be drawn from a Poisson distribution with mean  $\langle k \rangle = B_* / B_{\rm{min}}$. We note that there will be infinitely many encounters as $B_{\rm{min}} \rightarrow 0$ so that it is in general not possible to sample the full distribution, but only its truncated form. However, this is sufficient, since encounters with very small shock parameters become irrelevant. In particular, we will show in Section \ref{sec:simulations} that the joint effect of $k$ encounters is more or less equivalent to a single encounter with an effective shock strength,
\begin{align}
    B_{\rm{eff}} &= \left(\sum_{i=1}^k B_i^p \right)^{1/p}, \label{eqn:Beff}
\end{align}
with $p = 1.2$.

We sample a large number of shock histories and show the strongest three encounters together with the distribution of $B_{\rm{eff}}$ in Figure \ref{fig:shockdistribution}. The distribution of $B_{\rm{eff}}$ is already reasonably well approximated when only considering shocks with $B > 0.1 B_*$ i.e. approximately the 10 strongest shocks. It is almost fully converged when considering all shocks with $B > 10^{-3} B_*$ i.e. the 1000 strongest shocks. Typically the effective shock parameter is not too much larger than the strongest shock. Its median lies at $4.5 B_*$ whereas the median of the strongest shock lies at $1.46 B_*$. However, the low-end tail is shifted upwards quite a bit so that there are almost no cases with $B_{\rm{eff}} < B_*$. The high-end tail is virtually identical to the distribution of the strongest shock.

\section{Shock Distribution for orbits in the Milky Way} \label{sec:milkyway}

In this Section we evaluate numerically the quantities that are relevant for describing the full distribution of shock parameters for a cusp that is orbiting in the Milky Way. We showed in the last section that, for a given orbit, the time integral of the stellar mass density along the cusp's trajectory,
\begin{align}
    \chi_* &= \int \rho_*(\myvec{x}(t)) \, \mathrm{d}t, \label{eqn:chistar} 
\end{align}
is sufficient to parameterise the full distribution of encounter shocks. Our aim is thus to infer realistic estimates of $\chi_*$ and of the corresponding characteristic shock parameter $B_* =  2 \pi G \chi*$.

We assume that the orbital distribution of cusps follows that of dark matter particles within the Milky Way's halo.

\subsection{Milky Way Potential}
For the baryonic components of the Milky Way we assume the prescriptions used in the Phat ELVIS simulations \citep{kelley_2019} at the specific time  $z=0$.  The observational parameters underlying these simulations were taken from \citet{mcmillan_2017} and \citet{blandhawthorn_2016}.

The stellar disk and the gas disk are each modelled through a superposition of three \citet{Miyamoto_1975} (hereafter MN) potentials as described by \citet{smith_2015}. The parameters of the MN potentials have been tuned to recreate the mass distribution of an exponential disk up to $1\%$ accuracy. For the stellar disk we use a total mass of $M_{\mathrm{d}} = \SI{4.1e10}{\msol}$ a scale radius $R_{\mathrm{d}} = \SI{2.5}{\kilo \parsec}$ and a height parameter of $z_{\mathrm{d}} = \SI{0.35}{\kilo \parsec}$. For the gas disk we use $M_{\mathrm{d}} = \SI{1.9e10}{\msol}$, $R_{\mathrm{d}} = \SI{7.0}{\kilo \parsec}$ and $z_{\rm{d}} = \SI{0.08}{\kilo \parsec}$.

We approximate the stellar bulge through a \citet{hernquist_1990} distribution with mass $M_{\rm{B}} = \SI{9e9}{\msol}$ and with scale-length $r_{\rm{a}} = \SI{500}{\parsec}$.

For the Milky Way's dark matter halo, we assume that in the absence of baryonic components it would correspond to an NFW halo with concentration $c=8.7$ and mass $m_{200c} = \SI{e12}{\msol}$. However, the baryons increase the dark matter density in the inner regions. We model this contraction using the semi-analytic approach presented in \citet{cautun_2020}. Previously, we used the analytic  procedure of \citet{sellwood_2005}, but this produces a slightly denser result in the central regions. We prefer the \citet{cautun_2020} approach, since it has been tested in detail both against state-of-the-art simulations and against observations of the Milky Way.

\begin{figure}
     \centering
     \includegraphics[width=\columnwidth]{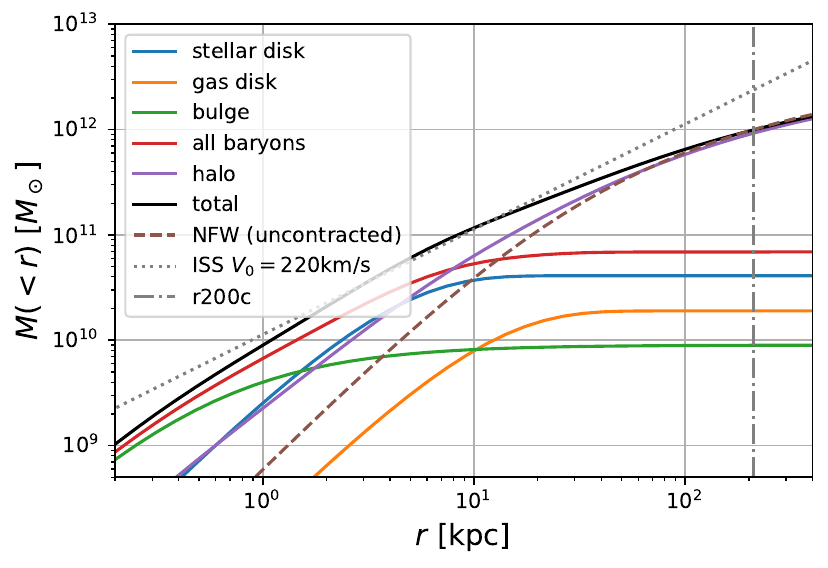}
     \caption{Enclosed mass profiles for the different components of our Milky Way model. Solid lines are used for components that contribute to the final potential. The brown dashed line shows the uncontracted NFW halo profile whereas the purple curve shows the profile after contraction due to the baryons and is the curve actually used in our modelling. The dotted grey line indicates the profile of a singular isothermal sphere with $V_0 = \SI{220}{\kilo \metre \per \second}$ which is sometimes used to approximate the spherically averaged potential of the Milky Way \citep[e.g.][]{errani_2021}.}
     \label{fig:enclosedmass}
\end{figure}

We show the result of the contraction of the halo together with profiles of all the different components in Figure \ref{fig:enclosedmass}. We note that the contracted halo (in purple) is significantly denser in the centre than the uncontracted version (brown dashed line). This contraction is only moderately relevant for getting the correct potential structure of the Milky Way. However, it is very relevant for sampling self-consistent orbits of the dark matter distribution. To sample orbits we use the \citep{Eddington1916} inversion method on the density profile of the contracted halo, but inside the joint potential of all components. We have checked that if we sample particles from the contracted halo profile and integrate their orbits in the joint potential, the density profile is stable and evolves very little at later times.

\subsection{Orbits}

We create $10^5$ test particles from the contracted halo up to \SI{400}{\kilo \parsec}. We use an adaptive sampling method so that their initial number density is proportional to $1/r$ times that of the halo. This way we have a good sampling, both at small radii $r \lesssim \SI{1}{kpc}$ and close to the virial radius $r \gtrsim \SI{200}{kpc}$. Throughout this paper, when  showing distributions,  we always correct this non-uniform sampling using the appropriate weights. We integrate particle orbits in the full 3d Milky Way potential (including bulge, stellar disk, gas disk and contracted halo) using $10^5$ constant timesteps over a period of $\SI{10}{\giga \year}$. Along each orbit we evaluate $\chi_*$ according to equation \eqref{eqn:chistar} using the stellar  density inferred from the Laplacian of the potential of the stellar components (stellar disk and bulge) $\rho_* = \Delta \phi_*  / (4 \pi G)$. Note that this can create negative densities in a (very) few locations, since the MN potential of the disk can create negative densities. We therefore clip $\chi_*$ after the integration at a minimum of $\SI{e-8}{\msol \per \parsec^2 \per (\kilo  \metre \per \second)}$. This threshold has no impact on any of our results. The actual minimum should be set by the diffuse stellar halo (and partially by encounters with dark substructures). However, none of these aspects matter since, as we will see in Section \ref{sec:smoothtides}, the effect of the smooth tidal field is much larger than this at radii where stellar encounters are infrequent.

\begin{figure}
    \centering
    \includegraphics[width=\columnwidth]{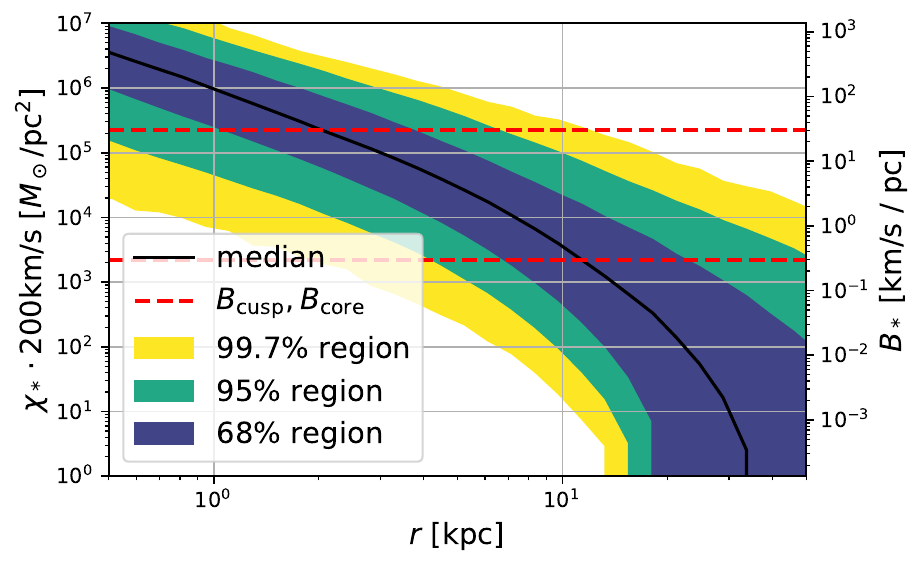}
    \caption{The distribution of $\chi_*$ as a function of radius. To facilitate interpretation, the left y-axis gives $\chi_*$ multiplied by a velocity of $\SI{200}{\kilo \metre \per \second}$; this estimates the total encountered stellar column density. An alternative y-axis shows the values of $B_*$, with red dashed lines showing the critical values, $B_{\rm{core}}$ and $B_{\rm{cusp}}$. It is clear that shocks will be relevant for almost all cusps that orbit in the central $\SI{10}{\kilo \parsec}$ of our Galaxy.}
    \label{fig:chistar}
\end{figure}

We find the distributions of $\chi_*$ and $B_*$ shown in Figure \ref{fig:chistar}. Here and in later plots we assign each particle 1000 times with its final value of $\chi_*$ at different radii chosen uniformly in time over its full orbit history. To help with an intuitive understanding of the $\chi_*$ distribution, we have multiplied the $\chi_*$ axis by a value of $\SI{200}{\kilo \metre \per \second}$ so that the left y-axis would correspond to the total stellar column density if the cusp always encountered stars at $\SI{200}{\kilo \metre \per \second}$. Figure \ref{fig:chistar} shows that a typical cusps orbiting around the solar radius (\SI{8}{\kilo \parsec}) encounters a stellar column density of about $\SI{e4}{\msol \parsec^{-2}}$. These numbers can easily be understood, as orbits at these radii will have passed through the Galactic disk about $10^2$ times and each disk passage adds a column density of order $\SI{e2}{\msol \parsec^{-2}}$. 

Further, we show as an alternative y-axis in Figure \ref{fig:chistar} the distribution of $B_*$. Recall that $B_*$ indicates the value of $B$ such that we expect on average one encounter with $B > B_*$. Therefore we expect typically one encounter with $B \gtrsim  \SI{1}{\kilo \metre \per \second \per \parsec}$ for cusps orbiting around the solar radius.

\begin{figure}
    \centering
    \includegraphics[width=\columnwidth]{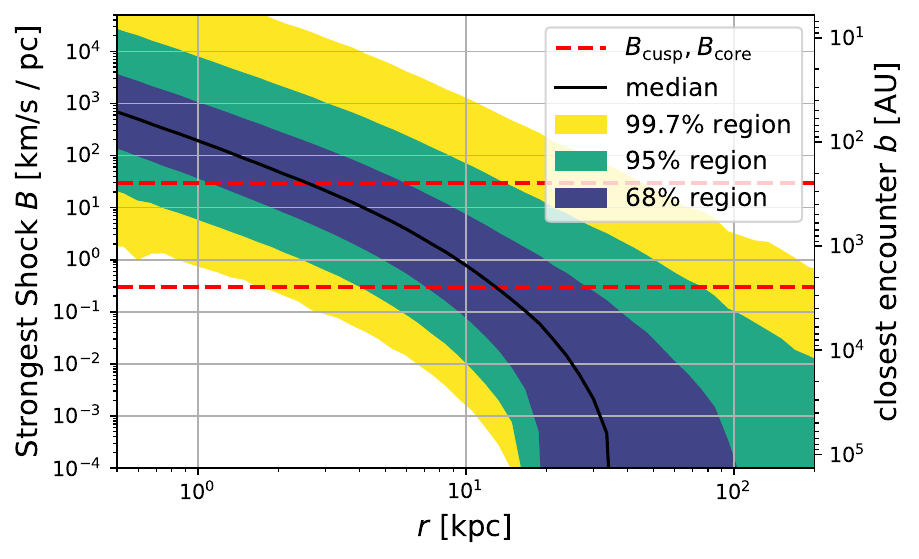}
    \caption{The distribution of the strongest shock encountered by a cusp as a function of its current Galactocentric radius. An alternative $y$-axis shows the corresponding closest approach distance for encounters with a mass of $\SI{1}{\msol}$ and velocity $\SI{200}{\kilo \metre \per \second}$. We see that within the central $\SI{10}{\kilo \parsec}$ the strongest shocks encountered will frequently  affect the annihilation signal of cusps and in some cases may disrupt them completely. }
    \label{fig:strongest_shock}
\end{figure}

For each orbit we sample the strongest encounters corresponding to the probability density in equation \eqref{eqn:strongestencounter}. In this way, we find the distribution of strongest shocks shown in Figure \ref{fig:strongest_shock}. We also show the corresponding encounter distance for encounters with mass $\SI{1}{\msol}$ and velocity $\SI{200}{\kilo \metre \per \second}$. This Figure is not used in our final results, but is useful to gain intuition for the relevant quantities.

We conclude that inside the central $\SI{10}{\kilo \parsec}$ virtually all cusps will have had at least one stellar encounter that might significantly decrease the annihilation luminosity and possibly even disrupt them. For a reliable evaluation, we have to conduct numerical experiments that evaluate how strongly cusps react to shocks of a given amplitude.

\section{The effect of impulsive encounters} \label{sec:simulations} 
To evaluate the effects of impulsive encounters on prompt cusps, we ran several sets of N-body simulations addressing scenarios of increasing complexity. The first set considers pure power-law profiles experiencing a single encounter. The second set considers cored profiles also with a single encounter. The third set considers both cored and power-law profiles experiencing multiple stellar encounters. For each case we develop simple descriptions for the annihilation luminosity expected from the remnants.

Finally, at the end of this section we briefly discuss how the effect of the smooth tidal field can be incorporated simply in this formalism.

\subsection{Simulation Setup}

We consider simulations starting from two different initial profiles. Power-law simulations start from a Poisson realization of a pure $r^{-3/2}$ density profile with the phase-space distribution from equation \eqref{eqn:plawphasespace}. In this case there is no relevant length or mass scale so that the results can be rescaled to any desired tidal shock strength. For cored simulations we use the phase-space distribution from equation \eqref{eqn:phasespacecored} and the density profile that is self-consistently created from this distribution, as described in Section \ref{sec:phasespacecores}. For these we set $r_{\rm{core}} = 1$ so that length scales are given in units of the core radius.

To enhance the dynamic range that can be resolved in these simulations and to minimize the effects of the (numerically required) outer truncation radius, we use a similar non-uniform mass sampling strategy to \citet{delos_2019_encounters}. Specifically, we arrange that the same number of particles $N_{\rm{split}}$ have pericentres in each of the radial ranges $(0, 1), (1, 10), (10, 100), (1000, 10000)$ which then have particle masses increasing by about a factor 30 between intervals. We explain this in more detail in Appendix \ref{app:nonuniformmass}, where we also present some numerical stability tests. An implementation of this method is available in the code repository of this article. We do not find any artifacts due to the non-uniform mass sampling, most likely because we use quite high resolution and because our pericentre criterion minimizes the intrusion of higher mass particles into higher resolution regions. 

For simulations with a single encounter (presented in Sections \ref{sec:powerlawtrunc} and \ref{sec:coredtrunc}) we apply at time $t = 0$ a single kick according to equation \eqref{eqn:kick} using the tensor from equation \eqref{eqn:kicktensor} with shock amplitude $B$. The shock introduces the characteristic spatial scale,
\begin{align}
    r_B &= \left(\frac{8 \pi G A}{3 B^2 }\right)^{2/3},
\end{align}
which is the radius where the tidal shock causes a (maximal) velocity change equal to the circular velocity in a pure power-law profile. By definition, $r_B$ equals $r_{\rm{cusp}}$ and $r_{\rm{core}}$ for shocks with strength $B_{\rm{cusp}}$ and $B_{\rm{core}}$ respectively, so that realistic shocks can easily produce $r_B$ values that lie at any point of the cusp (compare Figure \ref{fig:strongest_shock}). Further, the kick introduces a characteristic time-scale -- corresponding to the dynamical time at this radius:
\begin{align}
    t_B &= \frac{r_{B}}{v_{\rm{circ}} (r_{\rm{B}})} = \frac{1}{B}.
\end{align}
After the initial shock we evolve the simulation for several dynamical times $t_B$ until the profile no longer evolves at the radii of interest (e.g. $\sim 10 t_B$ around $r_B$). Note that for typical shocks with $B \sim \SI{1}{\kilo \metre \per \second \per \parsec}$, we have $t_B \sim \SI{1}{\mega \year}$.

\subsection{Truncation of Pure Power-law Profiles} \label{sec:powerlawtrunc}

\begin{figure}
    \centering
    \includegraphics[width=\columnwidth]{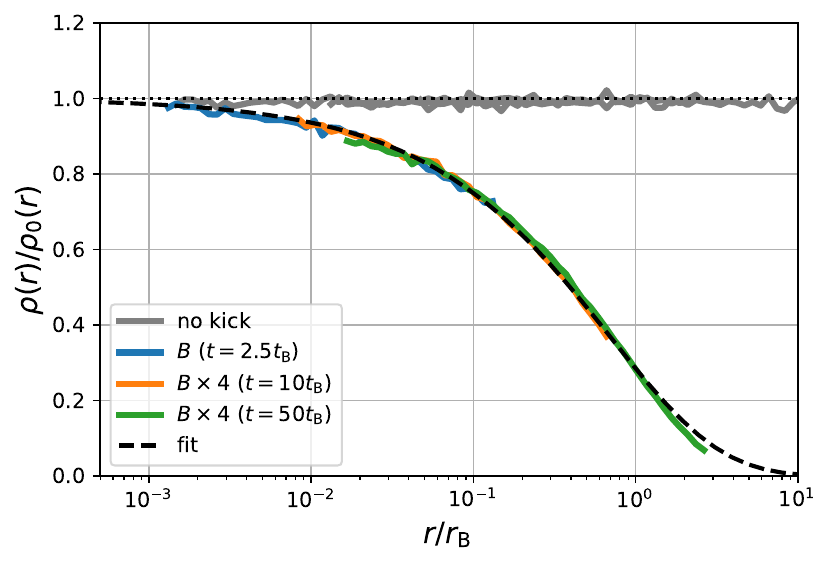}
    \caption{The transfer function for pure $r^{-3/2}$ power-law profiles exposed to a tidal shock from a stellar encounter. When presented in units of $r_B(B)$ the transfer functions of simulations with different shock amplitudes (blue versus orange and green) line up perfectly -- as is expected from dimensional analysis. The grey lines show the transfer functions of equivalent simulations without tidal shocks evaluated at the same output times -- proving the numerical stability of our setup. The black line shows the fit to the transfer function given in equation \eqref{eqn:powertransfer}. }
    \label{fig:powerlaw_transfer}
\end{figure}

We ran two high-resolution simulations of shocked power-law profiles which each have $N_{\rm{split}} = 2^{22}$ particles per radial interval, so that they have in total about $\SI{2e7}{}$ particles. These two simulations used different amplitudes for the shock parameter, varying by a factor of 4, leading to differing shock radii, $r_B = 20$ and $r_B = 128$. We remind the reader that, in principle, the result of a shocked power-law can be rescaled arbitrarily, so these two simulations differ only in how the physical scale $r_B$ compares to resolution parameters.

We define the transfer function,
\begin{align}
    T(r) &= \frac{\rho(r)}{\rho_0(r)},
\end{align}
where $\rho(r)$ and $\rho_0(r)$ are the final and initial density profiles, respectively, and we present transfer functions for our simulations in Figure \ref{fig:powerlaw_transfer} as functions of $r/r_{B}$. We show only the radial range that has reliably converged to a final, stable post-shock profile, and we use different output times to probe different parts of the transfer function. In Appendix \ref{app:convergence} we provide convergence tests to determine these scales. At the small end this is the largest radius where two-body relaxation is irrelevant, while at the large end we limit to scales where at least ten dynamical time-scales have passed. As additional evidence that the simulations are converged, we also show as grey lines in Figure \ref{fig:powerlaw_transfer} reference simulations evolved to the same times but with no shock. Clearly, these are stable and show no discernable numerical evolution. 

The simulations of different shock strengths line up very well if radii are scaled by $r_B$. In turn, since $B \propto b^{-2}$, $r_B$ scales with impact parameter as $b^{8/3}$. This is very different than the scaling of $b^{8/11}$ which \citet{Ishiyama_2010} assumed (without further explanation) to extrapolate their results and to argue that stellar encounters should be irrelevant for the survival of cusps. Such a scaling is clearly incorrect and is also inconsistent with the profiles of the simulations that \citet{Ishiyama_2010} presented themselves. These simulations, in fact,  seem consistent with a $b^{8/3}$ scaling and agree qualitatively with what we find here. 

Inspired by \citet{delos_2019_encounters}, we fit the simulated transfer functions jointly using a function of the form,
\begin{align}
    T(r) &= \exp(-\alpha (r/r_B)^\beta), \label{eqn:powertransfer}
\end{align}
where we find $\alpha = 1.256$ and $\beta = 0.639$ as the best fitting parameters. Interestingly the value of $\beta$ that we find here for our power-law profile is not too far from the one  that \citet{delos_2019_encounters} found ($\beta = 0.78$) for an NFW initial structure. The small difference presumably reflects the different central slopes, $-1.5$ and $-1$. 

We note that our measured profiles disagree slightly with our fit at larger radii $r \gtrsim 2 r_B$. However, this is quite irrelevant for the calculation of the annihilation luminosity where those contributions are downweighted by a factor $T^2(r)$.  For our fitted function we find that the annihilation rate is given by
\begin{align}
  J_{\rm{pow}}(B) &= 4\pi \int_{r_{\rm{min}}}^{r_{\rm{cusp}}} T(r)^2\rho_0^2(r) r^2  \,\mathrm{d}r \nonumber \\
          &= \frac{4 \pi A^{2}}{\beta} \left[\operatorname{Ei}{\left(- 2 \alpha \left(\frac{r_{\rm{min}}}{r_B}\right)^\beta \right)} - \operatorname{Ei}{\left(- 2 \alpha \left(\frac{r_{\rm{cusp}}}{r_B}\right)^\beta \right)}\right] \label{eqn:JtransferBoundary}  \\
          &\approx \frac{4 \pi A^{2}}{\beta} \operatorname{Ei}{\left(- 2 \alpha \left(\frac{r_{\rm{min}}}{r_B}\right)^\beta \right)}, \nonumber
\end{align}
 where $\operatorname{Ei}$ is the exponential integral and $r_{\rm{min}}$ is a lower limit of integration that is necessary to obtain a finite result. The approximation that is used in the last line,  $r_{\rm{cusp}} \rightarrow \infty$, is already very accurate for $r_{\rm{cusp}} \gtrsim r_B$. Therefore, if $r_B \leq r_{\rm{cusp}}$, it is fine to neglect the initial boundary of the system, since the truncation through the shock sets the actual boundary. Further it is insightful to consider the limit $r_{\rm{min}} \ll r_{\rm{B}}$ in which case
\begin{align}
  J_{\rm{pow}}(B)  &\approx \frac{4 \pi A^{2}}{\beta} \left(-\gamma - \log \left(2 \alpha \left(\frac{r_{\rm{min}}}{r_B}\right)^\beta  \right) \right)  \nonumber \\
          &= 4 \pi A^2 \left(\log \left(\frac{r_B}{r_{\rm{min}}} \right) -2.35 \right)  \nonumber \\
          &= 4 \pi A^2 \log \left(\frac{0.096 r_B}{r_{\rm{min}}} \right), \label{eqn:jpowerlawapprox}
\end{align}
where $\gamma \approx 0.577$ is the  Euler–Mascheroni constant. This approximation holds (at 5\% accuracy or better) for radii $r_{\rm{min}} < 0.1 r_{\rm{B}}$. One way to interpret this result is that the annihilation rate of the tidally truncated power-law corresponds to the annihilation rate of a power-law profile that is sharply truncated at $10\%$ of $r_B$ (compare Equation \eqref{eq:j_powerlaw}). This result is easily understood, since this is approximately the radius where $T^2$ reaches $50\%$. We expect this approximation to be inaccurate if $r_{\rm{core}} \gtrsim 0.1 r_B$, since then shock effects are significant at the core radius where they may be amplified. We will consider such cases in the next subsection and derive a general formula for the annihilation rate of shocked cusps. However, we can make a rough estimate of the cored annihilation rate, by assuming the power-law + transfer profile up to the core radius and down-weighting the core contribution by a factor of $T^2(r_{\rm{core}})$
\begin{align}
    J_{\rm{pow+core}}(B) &=  4 \pi A^{2}  \left(\beta^{-1} \operatorname{Ei}{\left(- 2 \alpha \left(\frac{r_{\rm{core}}}{r_B}\right)^\beta \right)}  + 0.531 T^2(r_{\rm{core}}) \right). \label{eqn:powerlawcoredann}
\end{align}
\subsection{Truncation of cored Profiles} \label{sec:coredtrunc}

\begin{table}
    \centering
    \begin{tabular}{c|c|c|c}
        $B / B_{\rm{core}}$ &  $r_{\rm{core}} / r_B$ & $t / t_B$ & $J / 4\pi A^2$ \\
        \hline
0.026 & 0.008 & 10 & 3.044 \\
0.053 & 0.020 & 20 & 2.213 \\
0.105 & 0.050 & 20 & 1.457 \\
0.211 & 0.125 & 20 & 0.723 \\
0.316 & 0.215 & 20 & 0.338 \\
0.421 & 0.316 & 20 & 0.134 \\
0.527 & 0.425 & 30 & 0.037 \\
0.632 & 0.543 & 50 & 0.003 \\
0.738 & 0.666 & 50 & 0.000
    \end{tabular}
    \caption{Simulation parameters for the shocked cored profiles: The shock parameter in units of the core scale, $B / B_{\rm{core}}$, the ratio of core to shock radius, $r_{\rm{core}} / r_B$, the duration of the simulation in units of the dynamical time, $t / t_B$ and the post-shock annihilation rate. The last simulation shows complete disruption.}
    \label{tab:cored_sims}
\end{table}

We also ran several simulations with cored initial conditions and with different shock parameters.  These are designed to probe the scales where the shock starts affecting the core. Each uses $N_{\rm{split}}=2^{20}$ and therefore in total about $\SI{5e6}{}$ particles. We evaluate each simulation at a time where the final profile has reached a stable result. This requires a longer integration time for simulations where the central density is reduced significantly.  We give the simulation parameters, durations and final annihilation luminosities in Table \ref{tab:cored_sims}.

\begin{figure}
    \centering
    \includegraphics[width=\columnwidth]{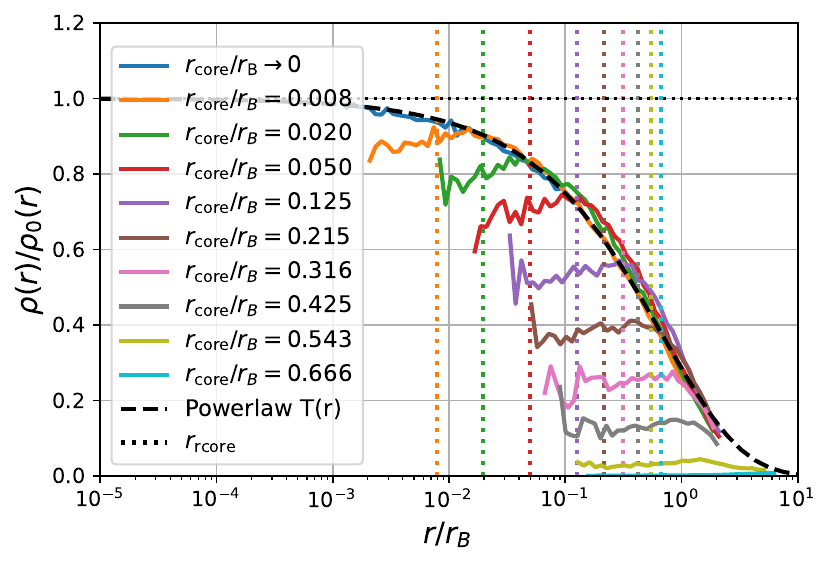}
    \caption{The transfer functions of cored power-law profiles. Note that each case has a different value of $r_{\rm{core}} / r_B$ so that the initial profiles are not identical as a function of $r/r_B$. Each simulation is divided by its own initial profile in making this plot. Core radii are indicated as vertical dotted lines. The transfer functions all show enhanced suppression relative to the power-law transfer function below and slightly above the core radius.}
    \label{fig:cored_transfer}
\end{figure}

We show the transfer functions of these simulations in Figure \ref{fig:cored_transfer}. We can see that as $r_{\rm{core}} / r_B \rightarrow 0$, the pure power-law behaviour is recovered; at large radii they have the same transfer function as the power-law case. At small radii cored profiles are suppressed additionally, and can even disrupt completely for sufficiently strong shocks. 

The simulation with $r_{\rm{core}}/r_B = 0.666$ is the first case which exhibits complete disruption -- this means that after $50 t_B$ no stable remnant is left, and densities keep decreasing everywhere. We checked that this case still disrupts if a four times higher particle number and a smaller softening are used, and that simulations with even larger shocks disrupt also. In addition, we a provide a video of one non-disrupting case and one disrupting case online.\footnote{\href{https://github.com/jstuecker/cusp-encounters}{https://github.com/jstuecker/cusp-encounters}}. It seems clear that although it is impossible to disrupt centrally divergent power-law profiles, it is indeed possible to disrupt cored profiles. This agrees with  theoretical \citep{amorisco_2021, stuecker_2022} and numerical investigations \citep{vandenbosch_2018, Errani2020, errani_2022} in the context of NFW haloes.

\begin{figure}
    \centering
    \includegraphics[width=\columnwidth]{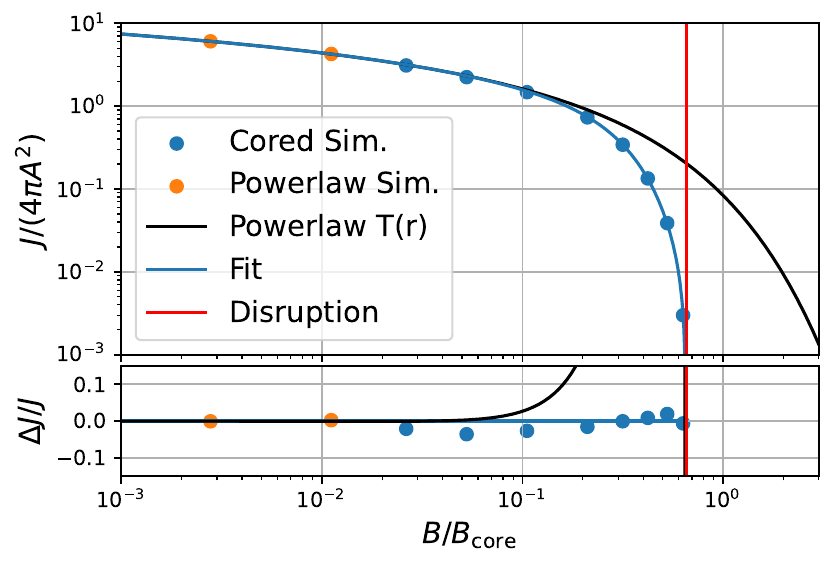}
    \caption{Annihilation rates of cored power-law profiles that have been exposed to shocks of varying amplitudes. The top panel shows the annihilation rates and the bottom panel the residuals with respect to our fit (blue line). The black line shows the estimate from equation \eqref{eqn:powerlawcoredann} based on the power-law transfer function. Clearly, the actual annihilation rates are additionally suppressed with respect to the power-law result when shocks get close to the core scale. For shocks with $B \gtrsim 0.65 B_{\rm{core}}$ we find complete disruption.}
    \label{fig:cored_annihilation}
\end{figure}

For the cored profiles we do not attempt to fit a functional form to the transfer functions, but instead directly calculate the annihilation radiation from the spherically averaged density profiles. This works very well numerically, because the part of the profile which is responsible for the bulk of the annihilation radiation (most significantly around $3 r_{\rm{core}}$) is resolved very well.\footnote{This is not the case  for our power-law profiles, where a major fraction of the radiation always comes from poorly resolved inner radii.} We describe the binning and integration procedures in Appendix \ref{app:annihilationconvergence} and show that all obtained annihilation luminosities are accurate at the $3\%$ level -- except for the last non-disrupted case, $r_{\rm{core}}/r_B = 0.543$, where the relative error is larger, but still less than $40\%$.

We list the annihilation rates we obtain in Table \ref{tab:cored_sims} and plot them in Figure \ref{fig:cored_annihilation}. In this Figure we also show the two power-law simulations with annihilation rates estimated by individually fitting their transfer functions and (arbitrarily) assuming a core radius, $r_{\rm{core}}=0.05$, for the annihilation estimate from equation \eqref{eqn:powerlawcoredann}. However, these two simulations could be arbitrarily rescaled along the black line.

We are able to fit the post-shock annihilation rates of our cored simulations using the function,
\begin{align}
    J &= 4\pi A^2 \left[0.531 +  \frac{1}{\beta} \log(a) - \frac{1}{\beta} \operatorname{Ei}\left(-2 \alpha a (x_{\rm{core}} + b x_{\rm{core}}^3)^{4\beta/3} \right)  \right], \label{eqn:Jcoredfit} \\
    x_{\rm{core}} &= \frac{B}{B_{\rm{core}}} = \left(\frac{r_{\rm{core}}}{r_B} \right)^{3/4},
\end{align}
with free parameters $a$ and $b$. By construction, this function approaches  the behaviour of the power-law truncation fit of equation \eqref{eqn:powerlawcoredann} for $B \ll B_{\rm{core}}$. However, it has two degrees of freedom $a$ and $b$ which can modify the behavior for large values of $B$.  
We find that $a=0.708$, $b = 5.98$ gives a reasonable fit to all annihilation rates at the $10 \%$ level -- both in the power-law regime $B \ll B_{\rm{core}}$ and when the shock hits the core $B \gtrsim 0.1 B_{\rm{core}}$. The function reaches zero at $B_{\rm{dis}} \approx 0.65 B_{\rm{core}}$.  
We assume that all cusps that had an encounter with $B > B_{\rm{dis}}$ are disrupted and have zero contribution to the annihilation rate. 

\subsection{The effect of multiple encounters} \label{sec:multikick}

As a final step, we need to understand what happens to the remnant if it is exposed to multiple shocks. To test this, we set up a large number of simulations (for both power-law and cored cases) with a variety of different shock histories. For each of these simulations we apply a shock approximately every ten dynamical time-scales $t_B$ and we use up to 5 shocks. We always compute the annihilation rate when $10 t_B$ have passed since the last shock. We note that the value of $t_B$ is not very clearly defined in the case with different shock amplitudes, we therefore choose a reference value $B_{\rm{ref}}$ to define a $t_B$ that seems appropriate for each individual history. We have checked that this shock interval is large enough to ensure that the results would not change by slightly decreasing it or by increasing it further. We list all shock histories in \ref{tab:shockhistories}. We have created several manually chosen shock histories (e.g. equal shock strengths, descending shocks, ascending shocks) and we have also sampled a few shock histories from the full distribution of shock histories with $B_* = B_{\rm{ref}}$, but only keeping the 5 strongest shocks -- see Section \ref{sec:shockhistories}. We sorted some of these histories accidentally, but we also created an additional set of four cored simulations with unsorted shock histories. However, neither the annihilation rate nor the transfer functions depend much on the order in which shocks happen.
\begin{table}
    \centering
    \begin{tabular}{c|c||c|c|c|c|c||c}
Type & $B_{\rm{ref}}$  & $B_1$ & $B_2$ & $B_3$ & $B_4$ & $B_5$ & $B_{\rm{eff}}$ \\
    \hline
          power-law &        - &        1 &        1 &        1 &        1 &        0 &      3.2\\
          power-law &        - &        1 &      0.5 &     0.25 &        0 &        0 &      1.5\\
          power-law &        - &     0.25 &      0.5 &        1 &        0 &        0 &      1.5\\
  power-law $B_*$, S &        - &      1.1 &     0.77 &     0.51 &     0.36 &     0.29 &      2.4\\
  power-law $B_*$, S &        - &      1.5 &     0.41 &     0.27 &     0.19 &     0.16 &      2.1\\
\hline
\hline
             cored &     0.11 &     1.00 &        1 &        1 &        1 &        1 &      3.8\\
             cored &     0.11 &     1.00 &      0.5 &     0.25 &        0 &        0 &      1.5\\
             cored &     0.11 &     0.25 &      0.5 &        1 &        0 &        0 &      1.5\\
     cored $B_*$, S &     0.11 &     1.08 &     0.77 &     0.51 &     0.36 &     0.29 &      2.4\\
     cored $B_*$, S &     0.11 &     1.45 &     0.41 &     0.27 &     0.19 &     0.16 &      2.1\\
     cored $B_*$, S &     0.11 &     9.05 &     0.79 &     0.68 &     0.67 &     0.43 &       10\\
     cored $B_*$, S &     0.11 &     1.97 &      1.6 &     0.65 &     0.46 &     0.42 &      4.0\\
     cored $B_*$, U &     0.05 &     3.43 &     0.34 &      1.4 &     0.28 &      1.6 &      5.7\\
     cored $B_*$, U &     0.11 &     1.03 &     0.57 &     0.42 &     0.65 &      4.1 &      5.6\\
     cored $B_*$, U &     0.11 &     0.35 &     0.38 &     0.44 &     0.18 &     0.28 &      1.3\\
     cored $B_*$, U &     0.26 &     0.96 &     0.15 &     0.89 &     0.28 &     0.23 &      2.0\\
    \end{tabular}
    \caption{The different shock histories simulated for the multiple encounters scenario. $B_{\rm{ref}}$ is given in units of $B_{\rm{core}}$ and all other shocks parameters are given in units of $B_{\rm{ref}}$. $B_1$ - $B_5$ indicate the strength of subsequent shocks and $B_{\rm{eff}}$ is the final value of the effective shock parameter, defined in equation~\eqref{eqn:effectiveshock} so that the effect of the whole shock history is equivalent to a single shock with this parameter. Shock histories that were sampled from the distribution of histories are indicated by a $B_*$. Those which were accidentally sorted in descending order are marked with an 'S' whereas unsorted ones are indicated by a 'U'.}
    \label{tab:shockhistories}
\end{table}

We present transfer functions for power-law simulations with multiple encounters in Appendix \ref{app:multikicktransfer}. Most importantly we note that cases with multiple encounters are still well approximated by the transfer function of equation \eqref{eqn:powertransfer}, but with different cut-off radii and we note that for the final transfer function, the order of shocks does not matter, only their amplitudes. These results are both consistent with findings in a previous study of NFW haloes \citep{delos_2019_encounters}.

\begin{figure}
    \centering
    \includegraphics[width=\columnwidth]{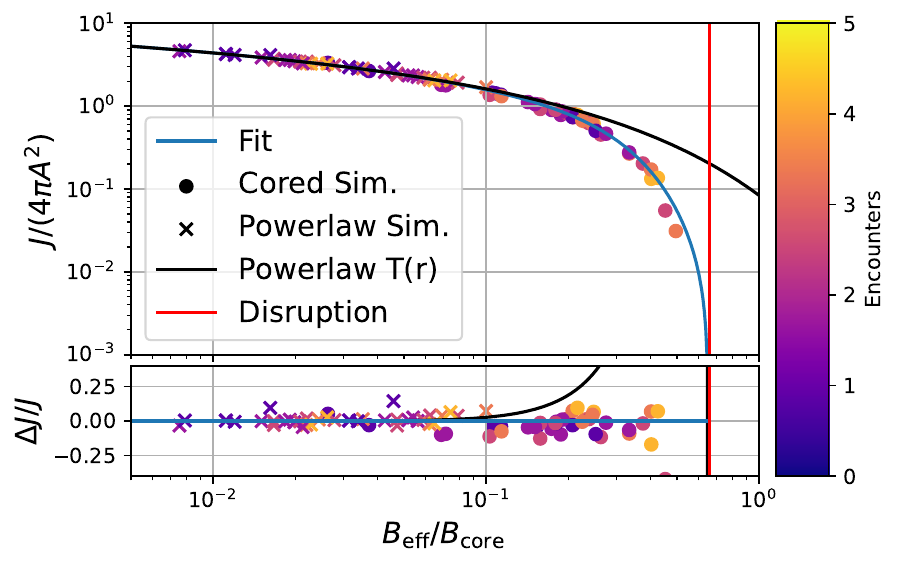}
    \caption{Annihilation rates for power-law and cored profiles after multiple stellar encounters. The encounter histories are summarized through a single effective parameter $B_{\rm{eff}}$ so that their annihilation rate is approximately equivalent to a single shock with $B_{\rm{eff}}$. The blue line shows our previous fit for single encounters, and the bottom panel shows residuals with respect to this fit. The effect of multiple encounters is captured to within 20\% through a single shock with the effective shock parameter $B_{\rm{eff}}$.}
    \label{fig:history_annihilation}
\end{figure}

We show the annihilation rates of cusps that have gone through multiple shocks in Figure \ref{fig:history_annihilation}. The data points in this Figure are obtained as follows. For power-law cases, we compute the annihilation rate by first fitting a transfer function to each encounter individually using equation \eqref{eqn:powertransfer} and then calculating the annihilation rate according to equation \eqref{eqn:powerlawcoredann}. Here we rescale the power-law results to two different core radii $r_{\rm{core}}=0.05$ and $r_{\rm{core}}=0.2$ so that each power-law simulation appears twice in Figure \ref{fig:history_annihilation}. We note again that these simulations could be rescaled to be anywhere along the black line. For cored profiles we calculate the annihilation rates as explained in the last section. For the $B$-axis we calculate an effective shock parameter which summarizes the whole history of encounters of a cusp through a single number,
\begin{align}
    B_{\rm{eff}} = \sqrt[p]{\sum B_i^p}, \label{eqn:effectiveshock}
\end{align}
which is the $p$-norm of the shock history. Different values of $p$ would give different importance to stronger versus weaker shocks. For $p = 1$ the value of $B_{\rm{eff}}$ would correspond to the sum of all shock parameters. For $p < 1$ multiple shocks would have an enhanced effect, and for $p \rightarrow \infty$ only the strongest shock would matter. For reference, we show, how such cases would appear in the Appendix \ref{app:multikicktransfer}. However, we have found that $p = 1.2$ gives excellent predictions, and this is the value of $p$ that we use for calculating the $B$-values in Figure \ref{fig:history_annihilation}. We note that some previous studies (of NFW subhaloes) have assumed that multiple shocks can be treated by adding the changes in binding energy \citep{shen_2022}. This would imply $p=2$ and would give clearly wrong results for the case of prompt cusps and probably also for NFW haloes. The results of \citet{delos_2019_encounters} indicate that multiple shocks have also a significantly enhanced effect ($p \ll 2$) for NFW haloes. However, it is not clear that our effective description will work equally well for NFW profiles, due to their more complicated form.   

The blue line in  \ref{fig:history_annihilation}   shows the prediction obtained by treating the shock histories as a single shock with effective shock parameter $B_{\rm{eff}}$ and inserting this into  Equation~(\ref{eqn:Jcoredfit})  This predicts the annihilation rates correctly to within $20 \%$\footnote{Only the last two data points have an error larger than this, but these points also have the largest systematic uncertainty, since they are so close to disruption. We would have needed to take additional care by choosing later evaluation times to get more precise estimates.}. We show in Appendix \ref{app:multikicktransfer} that this is much more accurate than results that would be obtained by only considering the strongest shock $p \rightarrow \infty$ (errors up to factor of a few) or by considering the sum of shocks as the effective shock parameter (errors up to $50\%$).

\subsection{Summary of the effect of stellar encounters}

We conclude that we can estimate the annihilation rate of cored power-law profiles that have gone through complicated shock histories simply by using equation \eqref{eqn:Jcoredfit} with an effective shock parameter calculated as the $1.2$-norm of all the shocks. To account for the initial boundary of cusps when $B \ll B_{\rm{cusp}}$ we additionally use the boundary term from equation \eqref{eqn:JtransferBoundary} so that we have
\begin{align}
    \begin{split}
    J = 4 \pi A^{2} \beta^{-1}  & \left[ - \operatorname{Ei}\left(-2 \alpha a (x_{\rm{core}} + b x_{\rm{core}}^3)^{4/3 \beta} \right)  \right. \label{eqn:Jcuspcore} \\
          &\left.  + \operatorname{Ei}{\left(- 2 \alpha x_{\rm{cusp}}^{4\beta/3} \right)} + 0.531\beta - \log(a)\right],    \\
    \end{split}
\end{align}
where we have defined
\begin{align}
    x_{\rm{cusp}} = \frac{B}{B_{\rm{cusp}}} = \left(\frac{r_B}{r_{\rm{cusp}}}\right)^{3/4}.
\end{align}
This is the main result of this section. Note that this function works accurately in all regimes -- it recovers the correct suppression for $B \gg B_{\rm{cusp}}$, but it also recovers equation \eqref{fig:cored_annihilation} for weak shocks with $B \ll B_{\rm{cusp}}$.

\begin{figure}
    \centering
    \includegraphics[width=\columnwidth]{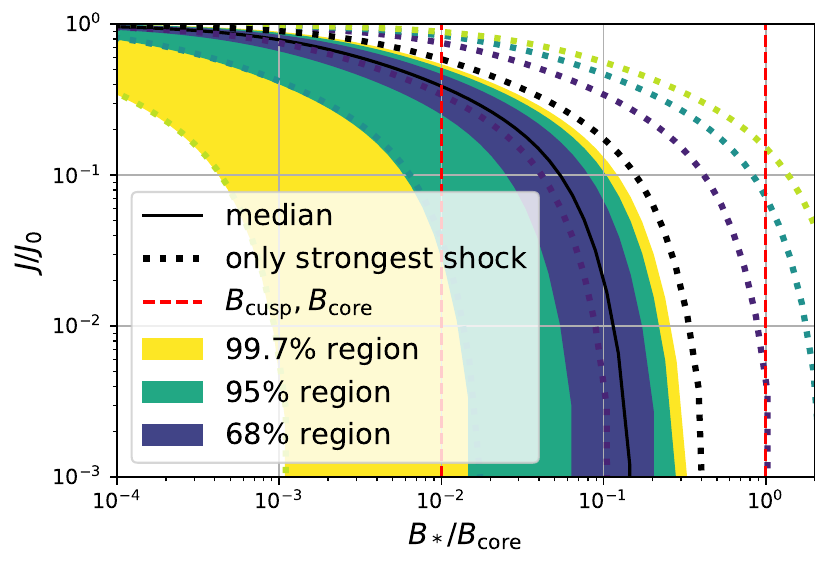}
    \caption{Reduction in annihilation luminosity due to encounters along a trajectory with a characteristic shock scale $B_*$ for a cusp with $B_{\rm{core}} = 100 B_{\rm{cusp}}$. The shaded regions show our predictions for the distribution of luminosities using the effective shock parameter. Dotted lines show the results that would have obtained by considering only the strongest shock. These have a slight offset in color so that they can still be seen where they overlap with the shaded regions. Clearly the effect of shocks with $B \gtrsim B_{\rm{cusp}}$ is quite dramatic, and when $B_*$ gets close to the core scale, complete disruption is expected in virtually all cases.}
    \label{fig:j_over_j0_vs_bstar}
\end{figure}

We note that the results we find here suggest that the impact of stellar encounters on the annihilation rate will be quite dramatic for cusps in the inner part of the Milky Way. We show in Figure \ref{fig:j_over_j0_vs_bstar} the distribution of the reduction in annihilation luminosity relative to the initial luminosity as a function of $B_*$, the characteristic shock strength of a cusp's trajectory. Here we assume $B_{\rm{core}} = 100 B_{\rm{cusp}}$, which is a typical ratio between these two parameters. We sample a large number of shock histories, considering all shocks with $B > 10^{-3} B_*$ and evaluate the expected luminosities according to \eqref{eqn:Jcuspcore} using the effective shock parameter. For comparison, we show also the luminosities that would be obtained by considering only the strongest shock. Virtually all cusps with $B_* \gtrsim 0.3 B_{\rm{core}}$ get completely disrupted and cusps with $B_* \gtrsim B_{\rm{cusp}}$ are already dramatically affected. As expected (compare also Appendix \ref{app:multikicktransfer}) there is a significant difference between considering only the strongest shock and considering the full history of shocks. However, even when considering only the strongest shock the suppression is quite strong.

\subsection{The effect of smooth tides} \label{sec:smoothtides}

In addition to stellar encounters, the smooth tidal field of the Milky Way can also induce mass-loss, and so a reduction in annihilation luminosity. We do not discuss the effect of smooth tides in great detail here, since this effect was already adequately incorporated by \citet{delos_white_annihilation_2022}, based on the work of \citet{stuecker_2022}. However, since the effect of smooth tides should be included in addition to that of stellar encounters, we need to make a few modifications to their approach. We will discuss these modifications in Appendix \ref{app:smoothtides}. In summary, we include the effect of smooth tides by applying the \textsc{adiabatic-tides} model \citep{stuecker_2022} to the cusps that remain after stellar shock truncation.  We find that the joint effect on annihilation radiation of a stellar shock with
effective strength $B_{\rm eff}$ and the smooth tidal field is approximately equivalent to a pure shock with an effective shock parameter,
\begin{align}
    B_{\rm{eff, \lambda}} &= \sqrt{B_{\rm eff}^2 + 42.2 \lambda}, \label{eqn:befflambda}
\end{align}
where $\lambda$ is the largest eigenvalue of the tidal tensor of the spherically averaged Galactic mass distribution at the  pericentre of the cusp's orbit.

\begin{figure}
    \centering
    \includegraphics[width=\columnwidth]{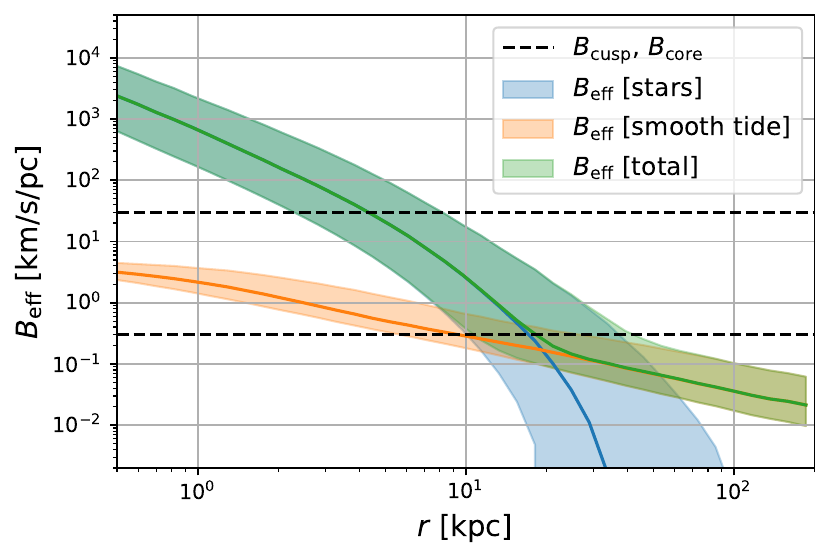}
    \caption{A comparison between the relative importance of the smooth tidal field and of tidal shocks from stars. The shaded regions indicate the $68\%$ regions of the distributions, and the solid lines their medians. For cusps that reach the vicinity of the disk, $r \leq \SI{20}{\kilo \parsec}$, encounters are dominant. In the outskirts of the Milky Way halo the effect of smooth tides dominates,
    but is too weak to affect the luminosity of most cusps.}
    \label{fig:beff_all_effects}
\end{figure}

In this way we are able to incorporate the effect of smooth tides simply through a redefinition of the shock parameter used in equation \eqref{eqn:Jcuspcore}. In Figure \ref{fig:beff_all_effects} we compare the effective shock parameter from the full history of shocks to the tidal contribution $B_{\lambda} = \sqrt{42.2 \lambda}$ and to the total effect. At radii $r \leq \SI{20}{\kilo \parsec}$ the effect of encounters dominates, whereas at larger radii the effect of the smooth tide dominates. Thus, viewed from the Earth's position just 8 kpc from the Galactic centre, truncation by stellar encounters has a large effect on the angular distribution of the prompt cusp annihilation signal. This must be taken into account when evaluating whether these cusps affect interpretation of the Galactic Center Excess measured by Fermi LAT. On the other hand, stellar encounters have much less effect on the radiation seen by a distant observer since most of the mass of the Milky Way's dark matter halo (and so most of its prompt cusps) are at larger radius.

\section{Results} \label{sec:results}

We combine results from Section \ref{sec:milkyway}, on the distribution of stellar  shock parameters, and from  Section \ref{sec:simulations}, on the effect of shocks and smooth tides on prompt cusps, to estimate the spatial distribution of the annihilation signal of cusps in the Milky Way.

For this we assume the cusp population expected  for a WIMP with mass  $\SI{100}{\giga \electronvolt}$ and decoupling temperature $\SI{30}{\giga \electronvolt}$. This leads to values $B_{\rm{core}}$ and $B_{\rm{cusp}}$ through equations \eqref{eqn:Bcusp_core} and to initial annihilation rates through equation \eqref{eqn:Jinitial}. We use the $10^5$ orbits and the final $B_*$ values inferred in Section \ref{sec:milkyway}, but we create a different realization of a cusp and its shock history for each of 1000 different radii sampled  uniformly in time along each orbit's trajectory. For the shock history we consider all shocks stronger than $B > 10^{-4} B_*$ (approximately the 10000 strongest shocks) when evaluating the effective shock parameter $B_{\rm{eff}}$. Additionally we keep track of the smallest radius $r_{\rm{peri}}$ that each cusp has reached and we evaluate the tidal field of the spherically averaged mass distribution  at this point to obtain the value of $\lambda$. With this we infer the effective shock parameter $B_{\rm{eff, \lambda}}$ as in equation \eqref{eqn:befflambda} and evaluate the expected final annihilation luminosity according to equation \eqref{eqn:Jcuspcore}. Thus, in total we obtain $10^{8}$ pairs of radii and annihilation luminosities.

\begin{figure}
    \centering
    \includegraphics[width=\columnwidth]{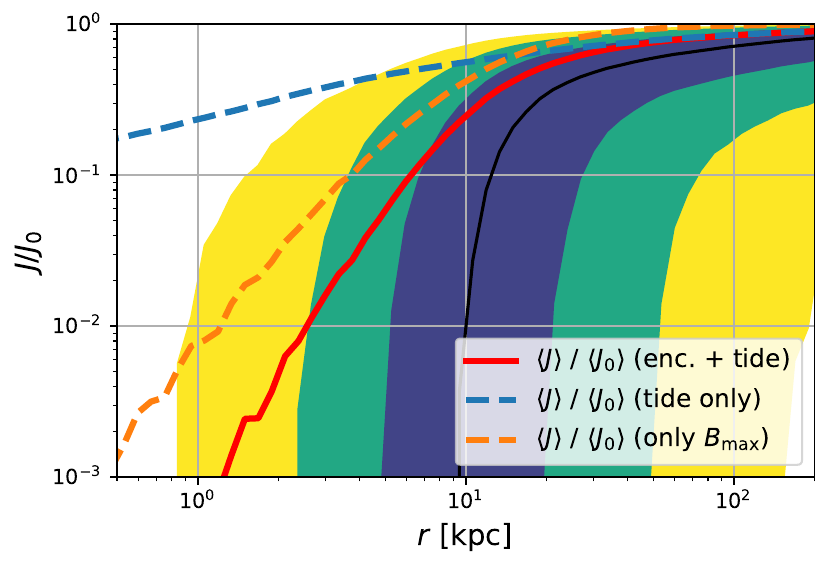}
    \caption{The distribution of the reduction in annihilation radiation from prompt cusps as a function of their current Galactocentric radius. Shaded regions indicate percentiles of the full distribution considering the joint effect of all encounters (colours as in Figure \ref{fig:chistar}). The mean (red line) gives the resulting effective reduction of the contribution to the annihilation profile from prompt cusps. The blue dashed line shows the mean reduction if only the smooth tide is considered, while the orange dashed line shows the reduction caused by  the strongest shock alone. Clearly, stellar encounters have a dramatic effect on the expected annihilation radiation from cusps in the inner Galaxy.}
    \label{fig:radial_annihilation_distribution}
\end{figure}

We show the distribution of the ratio between initial and final luminosities in the top panel of Figure \ref{fig:radial_annihilation_distribution} as a function of radius. The percentiles of this distribution give an idea of how dramatic the effect of stellar encounters on cusps is. Typical cusps inside of the central $\SI{10}{\kilo \parsec}$ are disrupted by stellar encounters. Within the central $\SI{3}{\kilo \parsec}$ less than $2.5\%$ of cusps survive, and almost all cusps reduce their luminosities by dramatic factors (e.g. $99.7\%$ of cusps reduce their luminosity at least by a factor 5). 

However, more relevant than the percentiles of the distribution is the annihilation weighted mean, since this gives the ratio between the initial annihilation profile and the final one. We show this as the red line in Figure \ref{fig:radial_annihilation_distribution}.  The mean is clearly dominated by the least disrupted cusps. This is so, since cusps that contribute more annihilation radiation are also more resilient to tides (compare Figure \ref{fig:cusp_B_distribution}) and further, since the bulk of the distribution gets completely disrupted, only the most resilient cusps with the least invasive shock histories contribute to the mean.  Even so, the mean is dramatically suppressed with respect to the case where encounters are neglected. At $\SI{8}{kpc}$ the average luminosity is already suppressed by a factor of 10. At smaller radii the effect is even more dramatic; only about $10^{-3}$ of the luminosity expected in the absence of encounters remains near the galactic bulge at about $\SI{1}{kpc}$. We show as a blue dashed line the mean annihilation reduction that would be obtained if only the effect of the smooth tide were considered \citep[as in][]{delos_white_annihilation_2022}. This greatly over-predicts the luminosity in the central regions, but is accurate at larger radii $r \gtrsim \SI{15}{\kilo \parsec}$. The orange dashed line shows the mean if only the strongest shock is considered instead of the full history. This leads to  significantly smaller, although still substantial, suppression.  

\begin{figure}
    \centering
    \includegraphics[width=\columnwidth]{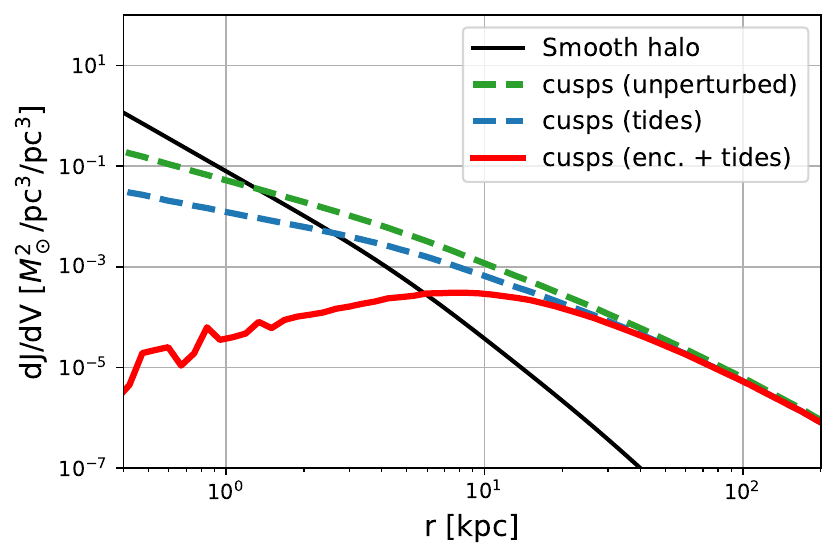}
    \caption{The radial distribution of annihilation radiation from cusps in the Milky Way including the effects of both stellar encounters and the mean tide (red line). For reference, the black line shows the radiation from the smooth halo, the green line the cusp contribution ignoring all disruptive effects, and the blue dashed line including  only the effect of smooth tides. When all effects are included the cusp contribution peaks at around 10~kpc.}
    \label{fig:radial_annihilation_profile}
\end{figure}

In Figure \ref{fig:radial_annihilation_profile} we show the annihilation profile as a function of radius. Unlike the prediction when only the effect of smooth tides is considered, stellar encounters lead to a non-monotonic profile that decreases towards the centre. Its maximum is slightly outside the solar radius at around $\SI{10}{\kilo \parsec}$.  At larger radii the effect of stellar encounters becomes irrelevant so that the original profile is recovered at $r \gtrsim \SI{40}{\kilo \parsec}$.

\begin{table}
    \centering
    \begin{tabular}{c|c|c|c|c}
        & initial   & only tides    & only enc. & tides + enc. \\
        \hline
      $J$-factor & 2.32e+11  & 1.90e+11  & 2.02e+11  & 1.79e+11 \\
       fraction & $100 \%$  & 82\%  &  87\% & 77\%
    \end{tabular}
    \caption{Total dark matter annihilation luminosity of the Milky Way when considering different tidal effects. $J$-factors are in units of $\SI{}{\msol \parsec^{-3}}$. These numbers are proportional to the luminosity of the Milky Way as  seen by a distant observer. Modelling tidal stripping or stellar encounters has only moderate effects on the total luminosity.}
    \label{tab:jfactors}
\end{table}

Let us briefly consider, how these effects alter the annihilation luminosity of the Milky Way, as seen by a distant observer. For this, we simply have to sum up the luminosity of all cusps out to some truncation radius which we assume to be  $r_{\rm{200b}} = \SI{340}{\kilo \parsec}$ (the radius where the enclosed mean density is 200 times the background density). Additionally we add the contribution of the smooth halo, which is however very small ($\sim 2\%$). We list the resulting total luminosities in Table~\ref{tab:jfactors}. Clearly, the joint effect of tidal stripping and stellar encounters onto the total luminosity is relatively small, only reducing the estimated luminosity by $23\%$. Stellar encounters do not much affect the total luminosity, since most of the cusps do not get close to the stars. As a result, it is not necessary to consider stellar and tidal disruption effects on cusps  when inferring the total contribution of extragalactic sources to the IGRB. However, we observe the dark matter distribution of our own Galaxy from a highly biased view-point  which greatly increases the sensitivity to what happens in its inner regions.

\begin{figure}
    \centering
    \includegraphics[width=\columnwidth]{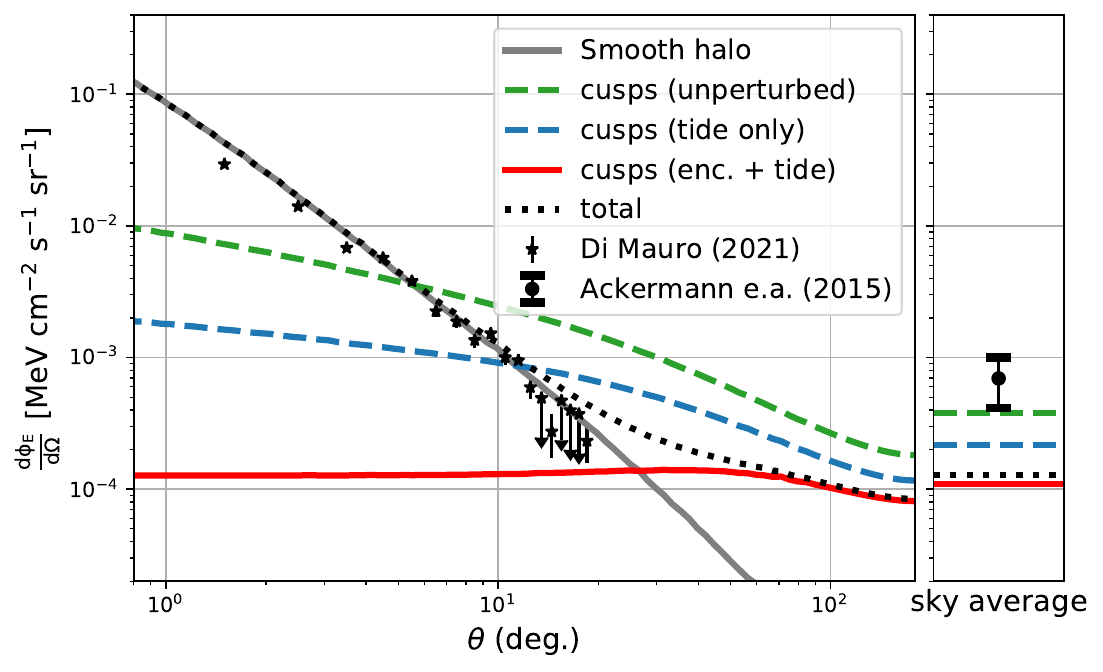}
    \caption{Left: The annihilation flux as a function of Galactocentric angle. When including the effect of stellar encounters the angular dependence of the prompt cusp contribution vanishes almost completely, resulting in a total signal which is compatible with the observed Galactic Centre excess \citep{di_mauro_2021}. Right: The sky average is significantly reduced due to stellar encounters, but cusps still contribute a significant and potentially  detectable fraction of the isotropic $\gamma$-ray background.}
    \label{fig:angle_annihilation_profile}
\end{figure}

We show in Figure \ref{fig:angle_annihilation_profile} the radiation profile of the Milky Way as it would be observed from the solar radius $r = \SI{8.2}{\kilo \parsec}$ if dark matter has a significant self-annihilation cross-section. Here we have inferred for each component the line-of-sight J-factor column-densities,
\begin{align}
    n_J &= \int_{0}^\infty \rho^2 \, \mathrm{d}l,
\end{align}
and then multiplied them by a normalization factor,
\begin{align}
    \frac{\mathrm{d} \Phi_E}{\mathrm{d} \Omega} = n_J \times \SI{1.44e-5}{\mega\electronvolt\centi\metre^{-2}sr^{-1} \second^{-1} \parsec^5 \msol^{-2}},
\end{align}
that makes the smooth halo profile agree with the Galactic Centre excess (this is the same factor that \cite{delos_white_annihilation_2022} have used). Additionally we have overplotted the data points of the observed GCE as given by \citet{di_mauro_2021}. Many error bars are too small to see and some data points are only upper limits indicated by arrows.\footnote{Our halo profile does not fit the GCE perfectly for the innermost angles. We did not fit the profile to the GCE, but simply use the one that we have also used for orbital modelling in Section \ref{sec:milkyway}.} Stellar encounters affect the profile so strongly that there is only a very small variation with observation angle left. While the unperturbed profile and the profile including tides alone both seem in tension with the observed shape of the GCE, stellar encounters alleviate this tension and the shape of the GCE is again compatible with dark matter as an explanation. 
The signal from cusps in the Galactic halo makes an almost isotropic contribution -- only deviating by about $30\%$ from its sky-average at its brightest angle ($40^\circ$).  When making conclusions from the {\it shape} of the signal profile, including the effects of stellar encounters is crucial.

\begin{table}
    \centering
    \begin{tabular}{c|c|c|c}
           component & $n_J$ & Flux & Fraction of IGRB \\
           \hline
         smooth halo & 1.3e+00 & 1.9e-05 &   2.7\%\\
    cusps (no tides) & 2.6e+01 & 3.8e-04 &  54.8\%\\
       cusps (tides) & 1.5e+01 & 2.2e-04 &  31.3\%\\
cusps (tides + enc.) & 7.6e+00 & 1.1e-04 &  15.8\%\\
          cusps (extragal.) & 1.2e+01 & 1.7e-04 &  25.0\%\\
          \hline
          total DM & 2.1e+01 & 3.0e-04 &  43.5\%\\
     total DM (previous) & 2.8e+01 & 4.1e-04 &  59.0\%\\
    \end{tabular}
    \caption{Average J-factor column densities (in units of $\SI{}{\msol^2 \parsec^{-5}}$) and sky-averaged fluxes of $\gamma$-ray radiation (in units of \SI{}{\mega\electronvolt~\centi\metre^{-2}sr^{-1} \second^{-1} })
    for different components  when normalizing fluxes so that the Galactic Centre excess would be explained by the smooth halo signal. Fluxes have been averaged over the sky with $\theta > 20^\circ$. If the GCE is due to dark matter, annihilation radiation  should constitute about $43\%$ of the isotropic $\gamma$-ray background.}
    \label{tab:fluxes}
\end{table}

However, if the GCE is due to dark matter we still expect a significant contribution to the $\gamma$-ray background from prompt cusps. We list the sky-averaged flux that we predict for each component in Table \ref{tab:fluxes}. It is most interesting to compare these numbers to the observed isotropic $\gamma$-ray background (IGRB)
\begin{align}
    \frac{\mathrm{d} \Phi_{\rm{IGRB}}}{\mathrm{d} \Omega} = 6.9^{+3.1}_{-2.8} \times \SI{e-4}{\mega\electronvolt\centi\metre^{-2}sr^{-1} \second^{-1} }
\end{align}
where the error indicates the systematic uncertainty due to foreground modelling which we have assumed to accumulate linearly when integrating the spectrum as measured by \citet{ackermann_2015}. \rev{Here, we only consider the integrated flux of the IGRB, but in principle the full spectral shape can be compared to the DM prediction.} The observed IGRB only considers contributions from galactic latitudes $b > 20^\circ$. To approximately mimic this, we have only included contributions from the Galactocentric angles larger than $\theta > 20^\circ$ in the sky-averages listed in Table \ref{tab:fluxes}. Additionally we have indicated what fraction of the IGRB each component would comprise. Finally, we have also listed the extra-galactic flux here which we estimate in the same manner as \citet{delos_white_annihilation_2022} -- neglecting the effect from tidal fields, which might reduce the number by around $20\%$ as indicated by Table \ref{tab:jfactors}. 

In comparison to the smooth tide prediction \citep[equivalent to][]{delos_white_annihilation_2022}, the predicted background signal due to cusps in our own Milky Way goes down by a factor $2.0$ and the signal is by a factor $3.5$ smaller than the unperturbed one. After taking this correction into account, the total dark matter signal (smooth halo + Milky Way cusps + extra-galactic cusps) is dominated by the extra-galactic signal and goes down by roughly one third. \citet{delos_white_annihilation_2022} argue that the morphology of the signal that is expected from annihilation from prompt cusps and from dark matter decay are essentially the same. Therefore, they use constraints on the decay of  dark matter from \citet{Blanco_2019} and rescale them to obtain constraints on the dark matter annihilation cross-section. These constraints are proportional to the predicted background, thus the upper limit on the cross-section has to be increased by about one third. However, we note that these constraints depend quite strongly on how well the astrophysical contributions to the diffuse $\gamma$-ray background are understood \citep{Blanco_2019} and the assessment of uncertainties has not been very rigorous so far. For example, the constraints by \citet{Blanco_2019} vary by a factor of about 4 simply by considering different assumptions about how correlated the error-bars are. Therefore, to infer reliable constraints it will necessary to reanalyse the IGRB with a more sophisticated treatment of the systematic and statistical uncertainties. 

Perhaps even more intriguing than the implications for constraining dark matter, the predicted contribution to the IGRB offers an independent test for the dark matter interpretation of the Galactic Centre excess. If the GCE is due to annihilation, then we predict an additional approximately isotropic $\gamma$-ray signal that would comprise about $40\%$ of the observed  1 to 10~GeV background for a $\SI{100}{\giga \electronvolt}$ WIMP. If we additionally consider that different WIMP models can lead to a factor $\sim 2$ variation in predicted cusp J-factors  \citep[consider][Figure 7 and Equation 2]{delos_white_annihilation_2022}, the total dark matter contribution should range between $20\%$ and $80\%$ of the observed IGRB. Given current uncertainties in the
(apparently dominant) contribution of star-forming galaxies and AGN to the observed signal, it is unclear, whether there is room for such a component. A careful reevaluation of these other contributions to the $\gamma$-ray background is clearly well motivated. \rev{The combination of total flux, the spatial morphology and the spectrum of the IGRB can be used simultaneously to constrain the existence of such an additional component.} Our work here can be used to infer templates for \rev{such an analysis.}

The detection or exclusion of this additional component would confirm or contradict the dark matter interpretation of the GCE, and so substantially  advance our understanding of dark matter itself. Firm exclusion could rule out an annihilation interpretation of the GCE, while robust detection would strongly support this interpretation, since it would be a remarkable coincidence to find the GCE and the additional IGRB contribution  at the right relative level yet due to different astrophysics.

\section{Discussion} \label{sec:conclusion}

In this article we have modeled the effect of stellar encounters on the expected dark matter annihilation signal from prompt cusps orbiting within the Milky Way's halo, presenting several advances in the treatment of such encounters. 

Firstly, we have developed a new method for inferring the full history of the impulsive shocks experienced by a dark substructure as it moves through the Galaxy. This method is both simpler and more general than previous approaches. While we have focused on prompt cusps in this article, our results on the distribution of shock histories (Sections 
 \ref{sec:theory} and \ref{sec:milkyway}) could be applied to conventional NFW subhaloes also, as long as the dominant shocks are in the distant-encounter regime (masses $\lesssim 1 M_\odot$).

Secondly, we have performed idealized N-body simulations to infer how stellar encounters affect prompt cusps. Here,  we have for the first time considered the phase-space core that any (otherwise) centrally divergent profile must exhibit. Only when this core is considered can a prompt cusp disrupt. Our simulations allow us to derive accurate formula to describe the structural effect of encounters with arbitrary combinations of shock history, core radius, truncation radius, characteristic density and smooth tidal field. As a result, we are able to account for the joint effect of smooth tides and of any number of stellar encounters along the entire trajectory of any cusp.

While our model is relatively complete and comprehensive, a few of its assumptions could still be improved. We have assumed a static Milky Way potential, and simply integrated cusp orbits for  $\SI{10}{\giga \year}$ within it. A more accurate treatment would consider evolution of the host potential and of the stellar population it contains, and would follow cusps from their initial formation and growth through their accretion onto precursor objects, and finally onto the Milky Way itself. While the early stages of this process remain quite uncertain, we believe that our procedure should be relatively accurate and should be conservative for effects at later times, since the great majority of Milky Way stars formed (approximately) {\it in situ} in the disc and the bulge, and most of them are less than $\SI{10}{\giga \year}$ old.

Another uncertain point is the precise profile of the phase-space core of prompt cusps. Here, we used a simple heuristic approach to obtain a stable profile that is consistent with the phase-space density constraint, but a rigorous investigation of the central profiles with N-body simulations would be desirable. However, we expect the results of our study to be quite robust to this uncertainty; annihilation rates depend relatively weakly on the precise \rev{radial scale and shape of the core because most cusps in the central region of the Galaxy have had encounters that put them far beyond the threshold for disruption. To significantly alter our predictions for the Galactic Centre would require increasing the resilience of cusps to shocks ($B_{\rm{core}}$) by 1-2 orders of magnitude, which is not plausible given the strict phase-space density constraint.}

 When we apply our modeling to the prompt cusp population in the Milky Way's halo, we find that stellar encounters have a dramatic effect on any cusps that enter the star-dominated regions  -- the vast majority get disrupted within the central $\SI{10}{\kilo \parsec}$, leading to a radial annihilation radiation profile that peaks near $\SI{10}{\kilo \parsec}$, dropping strongly at smaller radii. While this has little effect on the luminosity of the Milky Way as seen by a distant observer ($\lesssim 20\%$), it strongly affects the annihilation flux observable from Earth. Stellar encounters destroy cusps so efficiently in the inner Galactic halo, that the surface brightness of cusp annihilation radiation is predicted to vary only slightly between the centre and anticentre directions. As a result, it has no noticeable effect on the Galactic Centre Excess, which therefore remains compatible with production by annihilation of the smooth dark matter distribution in the inner few kpc. This removes the issue raised by \citet{delos_white_annihilation_2022} who included cusp disruption due to the smooth Galactic tide but not that due to stellar encounters, and hence found a total annihilation profile apparently incompatible with the GCE profile of \citet[][see Figure~\ref{fig:radial_annihilation_distribution}]{di_mauro_2021}.

This opens up an intriguing new possibility. If the GCE is indeed due to dark matter annihilation, this implies an approximately isotropic annihilation signal from prompt cusps that would have an amplitude in the range $20\% - 80\%$ of the observed isotropic $\gamma$-ray background, depending on the mass and decoupling temperature of the WIMP. The results of \citet{Blanco_2019} suggest that a signal of this amplitude is inconsistent with the observed IGRB, since the latter appears to be explained entirely by emission from star-forming galaxies and AGN. We think, however, that our current results warrant a careful reevaluation of those of \citet{Blanco_2019} since it seems conceivable that some of the fitted templates might absorb a near-isotropic dark matter annihilation signal, or that the statistical treatment of systematic errors is not yet robust enough to exclude such a signal. If such a signal is firmly ruled out, it becomes unlikely that the GCE can be ascribed to annihilation radiation. On the other hand, if it were robustly detected, this would confirm the annihilation interpretation of the GCE.

\section*{Acknowledgements}

JS thanks Sten Delos for answering quickly and clearly all questions regarding the distribution and properties of cusps. JS thanks Mattia Di Mauro for providing data related to the GCE.
JS and RA thank all members of the cosmology group at Donostia International Physics Center for daily discussions and for the motivating research environment. JS and RA acknowledge  the  support  of the European  Research Council through grant number ERC-StG/716151. GO was supported by the National Key Research and Development Program of China (No. 2022YFA1602903) and the Fundamental Research Fund for Chinese Central Universities (No. 226-2022-00216).


\section*{Data Availability}

The code used to generate all results of this study other than the simulation-based analysis of Section \ref{sec:simulations} is available in an online repository under \href{https://github.com/jstuecker/cusp-encounters}{https://github.com/jstuecker/cusp-encounters}. Some results are based on the \textsc{adiabatic-tides} code which is also publicly available under \href{https://github.com/jstuecker/adiabatic-tides}{https://github.com/jstuecker/adiabatic-tides}. The simulation data from Section \ref{sec:simulations} will be shared on reasonable request to the corresponding author.
 



\bibliographystyle{mnras}
\bibliography{example} 




\appendix

\section{Numerical convergence}
\subsection{Non-uniform mass initial conditions} \label{app:nonuniformmass}
To simulate the behaviour of our cusps we have to truncate them at some radius $r_{\rm{max}}$. We first tried to run numerical experiments with uniform mass sampling and a sharp truncation radius. With particle numbers of order $10^{6}$ one can typically resolve scales down to $10^{-3}$-$10^{-2}$ $r_{\rm{max}}$ before two-body relaxation effects become substantial. However, at the same time a sharply truncated $r^{-3/2}$ power-law is far from equilibrium around the truncation radius. We found that truncation  still has significant numerical effects below $10^{-1} r_{\rm{max}}$ (depending on how many dynamical times are simulated). This behaviour is much worse than for profiles that are steeper in their outer regions, such as Hernquist or NFW profiles. As a result, only a small range of radii is reliably resolved and a lot of care has to be taken to get numerically converged results over a substantial radial range. Unfortunately, the transfer function
we wish to measure spans several orders of magnitude in radius.

We have therefore decided to abolish the uniform mass sampling and instead divide the profile into several populations of particles with differing mass. This allows us to get reliably converged results over a dynamical range of $10^4$ in radius.

We define five populations of particles where the $i$th population is defined so that all of its particles have their orbital pericenters in the range,
\begin{align}
    r_{\rm{peri}} \in [r_{\rm{split},i}, r_{\rm{split},i+1}],
\end{align}
and we choose $\myvec{r}_{\rm{split}} = (0, 1, 10, 100, 1000, 10000)^T$.

In each interval we choose to use the same number of particles $N_{\rm{split}}$. In principle one can create such a realisation by setting up several full profiles at different resolutions and then discarding all particles that do not fulfill the criterion of each population. However, this would be very inefficient for generating the innermost populations.

Instead, we can directly sample particles from the cumulative energy, pericentre distribution function. The density contribution at radius $r$ from particles which have pericentres in a radial range $r_1 < r_{\rm{peri}} < r_2$ and energy smaller than $E$ is given by
\begin{align}
    \rho(r, &r_1 < r_{\rm{peri}} < r_2, < E) \nonumber \\
            &= 4\pi \int_{\phi_*}^E \int_{L_1}^{L_2}  \frac{L f(E)}{r^2 \sqrt{2E - 2\phi - \frac{L^2}{r^2}}} \, \mathrm{d}L \, \mathrm{d}E,
\end{align}
where the boundaries $L_1$ and $L_2$ are chosen so that the integral goes only over particles which fulfill the pericentre criterion:
\begin{align}
    L_1 &=  \rm{max}(r_1 \sqrt{2(E - \phi(r_1))}, 0), \\
    L_2 &=  \rm{min}(r_2 \sqrt{2(E - \phi(r_2))}, r \sqrt{2(E - \phi(r))}).
\end{align}
We find that

\begin{align}
    \rho(r, &r_1 < r_{\rm{peri}} < r_2, < E) \nonumber \\
            &= F(E, \phi_{*1}) \sqrt{1 - \frac{r_1^2}{r^2}} - F(E, \phi_{*2}) \sqrt{1 - \frac{r_2^2}{r^2} }, \\
    \phi_{*1}(r) &= \frac{\phi(r) - \phi(r_1) \frac{r_1^2}{r^2}}{1 - \frac{r_1^2}{r^2}}, \\
    \phi_{*2}(r) &= \frac{\phi(r) - \phi(r_2) \frac{r_2^2}{r^2}}{1 - \frac{r_2^2}{r^2}},
\end{align}
with
\begin{align}
\begin{split}
    F(E, \phi) = \frac{8 \sqrt{2} \pi f_0 \left(E - \phi\right)^{\frac{3}{2}} }{105 \left(E + E_{c}\right)^{\frac{7}{2}} \left(E_{c} + \phi\right)^{3}} & \left( 8 E^{2} + 28 E E_{c} + 12 E \phi \right. \\
      & \,\, \left.+ 35 E_{c}^{2} + 42 E_{c} \phi + 15 \phi^{2} \right),
\end{split}
\end{align}
where the terms containing $r_1$ have to be set to zero if $r < r_1$ and the terms containing $r_2$ have to be set to zero for $r < r_2$.

Knowledge of this density function is enough to sample radii and energies of particles in each group. Radii can be sampled through inverse-distribution function sampling with the cumulative mass profile that can be obtained by integrating $\rho$ and using the limit $E \rightarrow \infty$, but maintaining a finite pericentre range. Energies can be sampled by inverse-distribution function sampling among energies using the normalized version of $\rho(r, r_1 < r_{\rm{peri}} < r_2, < E)$. Finally, angular momenta can be sampled by considering the cumulative angular momentum distribution function,
\begin{align}
    F(<L | r,E) &= \left[ \sqrt{2E - 2\phi - L^2/r^2} \right]_{L_1}^{L_2}.
\end{align}
An implementation of this scheme for creating initial conditions for cored cusps is also publicly available in the code repository of this paper.

\begin{figure}
    \centering
    \includegraphics[width=\columnwidth]{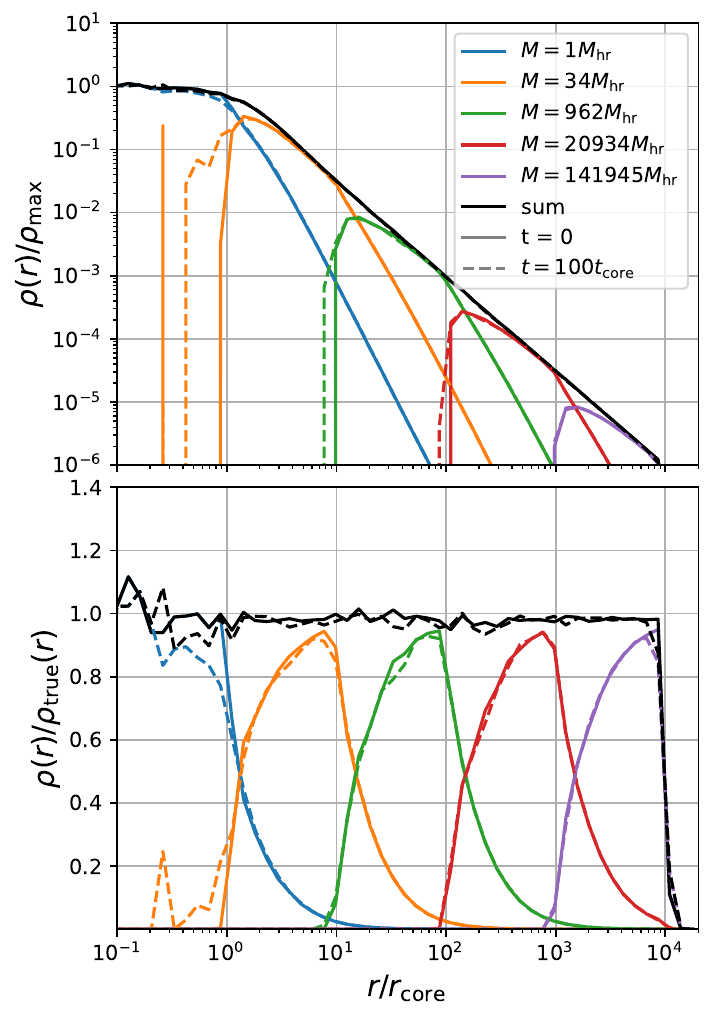}
    \caption{A cored density profile that has been set up through our non-uniform mass sampling technique (top) and the ratio to the true profile (bottom). Black lines show the full profile whereas colored lines show the individual components made of particles with different masses that were separated by their initial pericentre radii. The legend indicates the particle mass of each population in units of the mass of the highest resolution particles. Dashed lines show the profiles after evolving the simulation for 100 dynamical times. Clearly our non-uniform mass sampling creates very stable profiles over a large dynamical range.}
    \label{fig:adaptivesampling}
\end{figure}

We show the radial density profiles of the initial conditions that have been created in this manner for a profile with $r_{\rm{core}} = 1$ and $N_{\rm{split}}=2^{18}$ as solid lines in Figure \ref{fig:adaptivesampling}. The individual components are shown as coloured lines and the sum is shown as the black lines. Clearly this gives an excellent initial representation of the density profile. The dashed lines in the same Figure show the density profile after the simulation has been evolved for 100 times the dynamical time-scale at the core radius. The final profile still agrees excellently with the initial one and we can see that particles of different masses have mixed very little. Note that the particle number here is still a factor $4-16$ lower than for the simulations that we actually use in the main text so that we can be fairly confident that all simulations are stable and well converged.

\subsection{Time evolution and convergence of shocked power-law profiles} \label{app:convergence}
\begin{figure}
    \centering
    \includegraphics[width=\columnwidth]{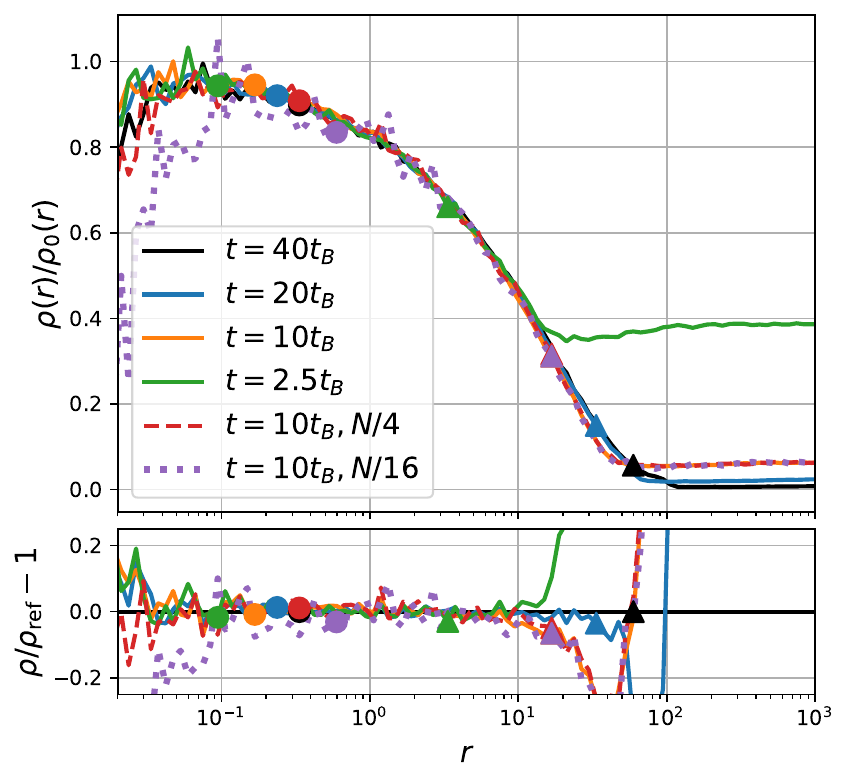}
    \caption{Convergence of the transfer function of a shocked power-law profile. The top panel shows the density transfer functions, and the bottom panel the residuals with respect to the $t = 40 t_{\rm{B}}$ reference case. Different lines indicate different times and/or simulations with different particle numbers.  The dots and triangles indicate the inner and outer convergence radii which are given on the inside by $r_{\rm{min}}$, a conservative estimate of the two-body relaxation radius, and on the outside by $r_{\rm{max}}$,  the radius where about 10 dynamical times have passed since the shock. (The outer orange and red triangles are overlayed by the purple one.) The simulations are very well converged inside this range and the inferred convergence radii are very conservative estimates.}
    \label{fig:transfer_convergence}
\end{figure}

We briefly discuss the convergence aspects of the shocked power-law profiles here. We show the profile of the $r_B = 20$ simulation with $N_{\rm{split}} = 2^{22}$ at different output times as solid lines in Figure \ref{fig:transfer_convergence}. The profile is very stable with time over large radial ranges, but there are differences at large radii where at early times the profile has not yet had enough time to relax to its final form. We find that a conservative criterion for the largest radius $r_{\rm{max}}$ where the profile has relaxed to its final form is given by
\begin{align}
    t_{\rm{sim}} &= 10 t_{\rm{dyn}} (r_{\rm{max}}) \nonumber \\
                 &= 10 \sqrt{\frac{r_{\rm{max}}^3}{M(<r_{\rm{max}}) G}},
\end{align}
where $t_{\rm{sim}}$ is the run-time of the simulation and we use the mass profile at the output time (not the initial time) for $M(<r)$. We invert this equation numerically to find $r_{\rm{max}}$ at each time and mark this point in Figure \ref{fig:transfer_convergence}. Clearly this is a very conservative criterion and cuts out all parts of the profile that are yet to reach their final form.

The innermost radius $r_{\rm{min}}$ where we can trust the simulations is given by the two-body relaxation radius which we can approximate through
\begin{align}
    t_{\rm{sim}} = t_{\rm{relax}}(r_{\rm{min}}) &\approx  \frac{0.1 N (< r_{\rm{min}})}{\log (r_{\rm{0}}/\epsilon)} t_{\rm{dyn}}(r_{\rm{min}})
\end{align}
\citep[e.g.][]{BinneyTremaine2008} where $r_{\rm{0}}$ is an estimate of the largest two-body encounter radius and $\epsilon$ is the softening. Since, the dependence on $r_{\rm{0}}$ is rather weak we just use $r_{\rm{0}} = r_{\rm{min}}$. This formula only holds strictly for systems that are made of uniform mass particles. Therefore, to be extra conservative, we use here not the actual particle number here, but the particle number as if all mass inside of $r_{\rm{min}}$ was made up from particles of the second highest resolution level (from the pericentre interval [1, 10]). Since most particles in the region are actually much lower mass (and none are higher mass), the profiles are actually converged at substantially smaller radii than the thus determined $r_{\rm{min}}$, which we indicate in Figure \ref{fig:transfer_convergence}. Further, we show two lower resolution simulations with $4$ and $16$ times fewer particles to demonstrate that the simulations are actually very well converged in the indicated regimes.  In Figure \ref{fig:powerlaw_transfer} we show only the reliable regions inside the range [$r_{\rm{min}}$, $r_{\rm{max}}$]. 

\subsection{Annihilation Rates} \label{app:annihilationconvergence}
We describe here how we compute the $J$-factors of the cored power-law profiles that we presented in Section \ref{sec:coredtrunc}. In many cases it is numerically problematic to directly infer annihilation rates from N-body simulations of centrally divergent profiles. This is so since annihilation rates scale as the density squared, and are therefore particularly sensitive to the central regions of the profile which are, however, in general also the regions that suffer the most from numerical inaccuracies caused by two-body relaxation, sampling noise and the  effects of force softening.

\begin{figure}
    \centering
    \includegraphics[width=\columnwidth]{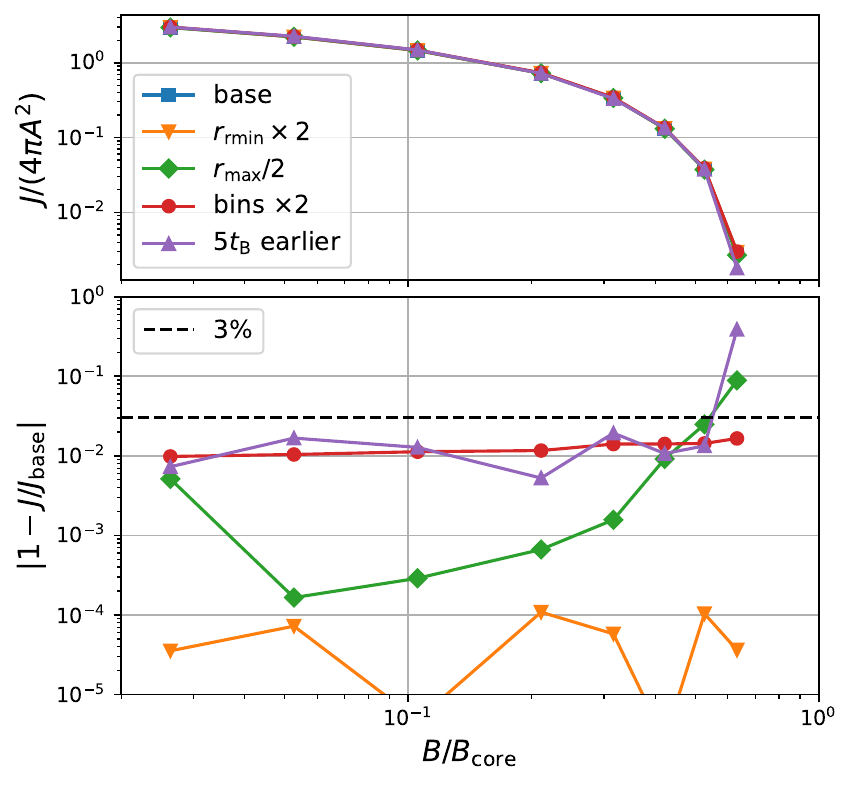}
    \caption{Convergence of the annihilation rates. The top panel shows the annihilation rates, whereas the bottom shows the difference relative to the reference case. Different lines show variations of numerical parameters with respect to the fiducial ones. All except the last data point have systematic errors well below $3\%$.}
    \label{fig:ann_convergence}
\end{figure}

However, for the simulated cored profiles none of these are problematic, since most of the annihilation radiation comes from radii $r \gtrsim r_{\rm{core}}$ which are resolved extremely well in our simulations. Therefore, we estimate the annihilation radiation in the following manner: We infer the radial density profile by assigning particles to 50 logarithmically spaced radial bins in the range [0.1 $r_{\rm{core}}$, $10^4 r_{\rm{core}}$] (i.e. 10 bins per dex). We then numerically integrate the $J$-factor by combining third-order spline interpolation with a large number of integration points which we place between $r_{\rm{min}} = 0.2 r_{\rm{core}}$ and $r_{\rm{max}} = 6 r_{B}$. Finally, we add the contribution from $r \lesssim r_{\rm{min}}$ by assuming that the density is uniform in this range with a value given by the average density within that radius. We show the resulting annihilation rates as the blue line in Figure \ref{fig:ann_convergence}. We then consider different variations of the numerical procedure: using a two times larger value of $r_{\rm{min}} = 0.4 r_{\rm{core}}$; using a smaller value for $r_{\rm{max}} = 3 r_{B}$; using two times the number of bins (20 per dex); and by evaluating the profile at a slightly different simulation time ($5 t_B$ earlier). Note that the last variation tests that the final profile is both stable and robust to shot noise. We show these cases in Figure \ref{fig:ann_convergence}, together with the residuals with respect to the reference case in the bottom panel. Clearly none of these deviations has a significant impact on the annihilation radiation. In all cases the associated relative error is less than $3\%$, except for the case of the strongest shock considered where the error is significantly larger, but still less than $40\%$. These results are more than accurate enough for the purposes of this paper.

\subsection{Transfer functions and multiple encounters} \label{app:multikicktransfer}
\begin{figure}
    \centering
    \includegraphics[width=\columnwidth]{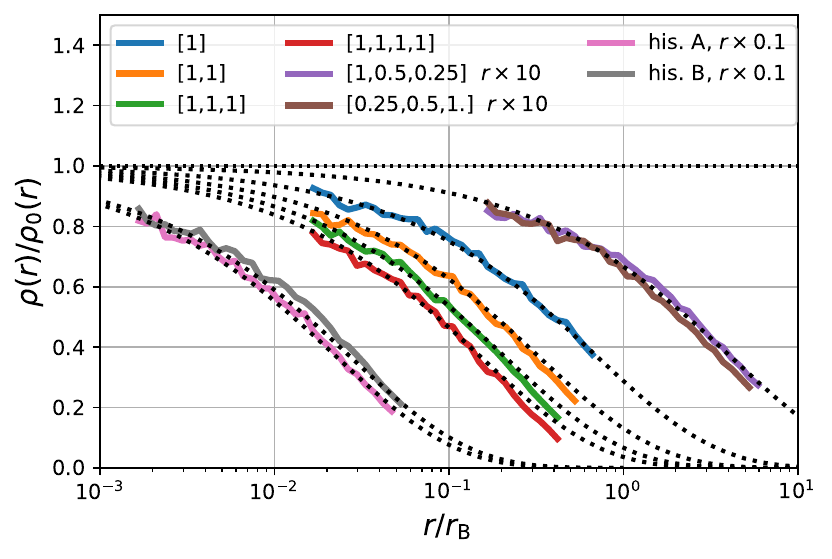}
    \caption{Transfer functions for power-law profiles after experiencing multiple encounters. Dotted lines indicate predictions using an effective shock parameter defined as the $p$-norm with $p = 1.2$. To avoid clutter, some lines have been offset by a factor 10 up or down in radius. For most cases the labels indicate the shock histories and for the last two cases the shock histories can be found in Table \ref{tab:shockhistories}.}
    \label{fig:multikick_transfer}
\end{figure}

Here, we present the transfer functions obtained for power-law profiles that have gone through up to 5 shocks in different sequences. Each shock is applied from a random direction. We use the power-law simulations that are listed in Table \ref{tab:shockhistories} and show their transfer functions in Figure \ref{fig:multikick_transfer}. Additionally, we show the transfer function that is predicted by treating the full history as a single encounter with effective shock parameter $B_{\rm{eff}}$, given by the $p=1.2$ norm of the shock history. Clearly this effective shock parameter predicts the transfer function of multiple encounters reasonably well.

\begin{figure}
    \centering
    \includegraphics[width=\columnwidth]{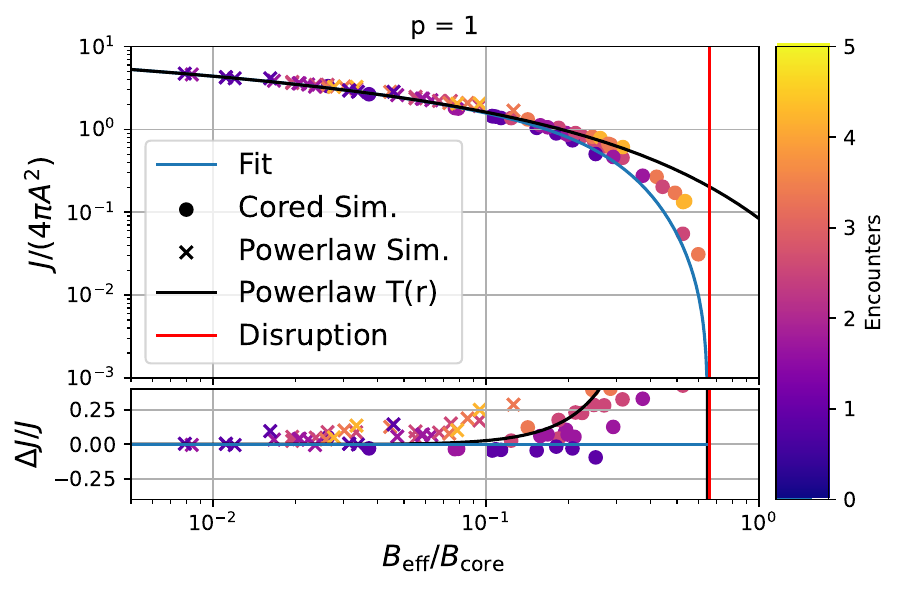}
    \includegraphics[width=\columnwidth]{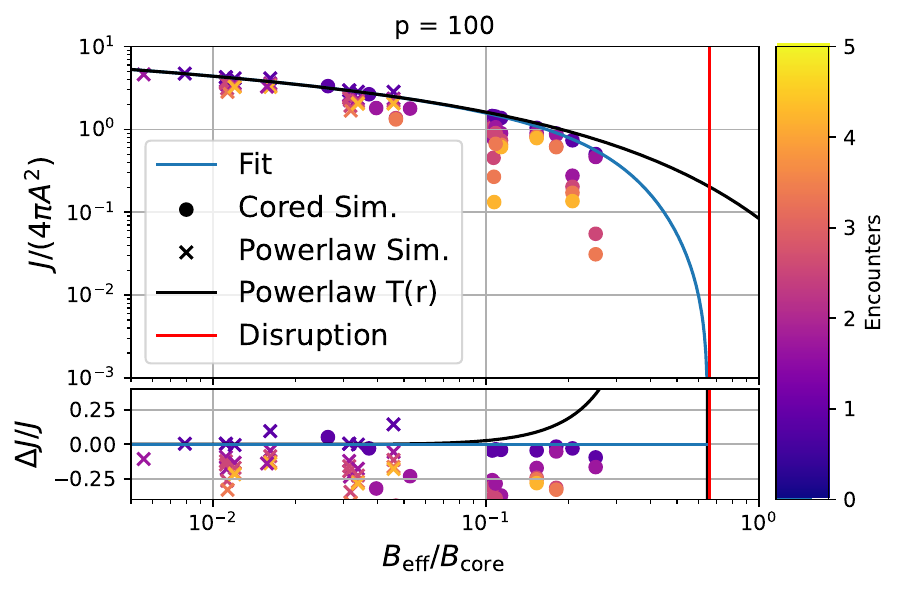}
    \caption{Multiple encounter annihilation luminosity predictions using different values of $p$ to calculate the $p$-norm of the shock history (for comparison with Figure \ref{fig:history_annihilation}) Top: $p=1$ corresponding to a linear sum of the shock parameters. This does not work well, giving  annihilation rates off by more than $50 \%$ for shocks with $B / B_{\rm{core}} \gtrsim 0.2$. Bottom: $p=100$ -- approximately corresponding to the $\infty$-norm, thus selecting only the strongest shock. This gives very poor predictions (off by factors up to $5$). It is clearly inadequate to consider only  the effect of the strongest shock.}
    \label{fig:J_different_p}
\end{figure}

Finally, we present alternative versions of Figure \ref{fig:history_annihilation} that use different values of $p$ for calculating the norm of the shock history. The top panel of Figure \ref{fig:J_different_p} assumes $p=1$, so that $B_{\rm{eff}}$ corresponds to the linear sum of $B$. In this case the prediction (blue line) overestimates the reduction in annihilation radiation by up to $50\%$. The bottom panel adopts the value $p=100$ -- effectively selecting only the strongest shock to define $B_{\rm{eff}}$. Clearly, considering only the strongest shock can dramatically underpredict the reduction in annihilation luminosity. For example,  at $B_{\rm{eff}} / B_{\rm{core}} \sim 0.1$ there are cases where the annihilation luminosity based on the full shock history is 10 times smaller than predicted from just  the strongest encounter. 

\section{The effect of the smooth tidal field} \label{app:smoothtides}
While the main text of this paper discusss the effect of stellar encounters in great detail, the smooth tidal field also affects the annihilation luminosity of prompt cusps orbiting in the Milky Way \citep[see e.g.][]{delos_white_annihilation_2022}. As discussed in \citet{stuecker_2022}, the most important parameter determining this effect is the largest eigenvalue $\lambda$ of the tidal tensor at the orbital pericentre of a prompt cusp's orbit. After a sufficient amount of time ($\gtrsim$ 10 orbits) the orbiting cusp approaches an asymptotic structure which changes little in subsequent evolution \citep{errani_2021}. \citet{stuecker_2022} show that the asymptotic remnant can be reasonably approximated by the \textsc{adiabatic-tides} model. This is an analytic description of an object's reaction when a tidal field is applied very slowly (in the adiabatic limit) and in a spherical approximation. The corresponding code can be found online.\footnote{\label{repository}\url{https://github.com/jstuecker/adiabatic-tides}} While actual N-body simulations in the Galactic context would of course be more accurate, the \textsc{adiabatic-tides} model allows easy exploration of a variety of different scenarios. We note that most studies of tidal stripping have focused on NFW subhaloes which are much less resilient to tides than prompt cusps. As a result, most previously published results cannot be applied to the case of interest here.

For a pure power-law profile with slope $-1.5$ the truncation radius due to smooth tides scales with a characteristic length $r_{\lambda}$, which may be defined by equating the cusp's attractive force and the disruptive force due to the tidal field:
\begin{align}
    \lambda \cdot r_{\lambda} &= \left| \frac{\partial \phi}{\partial r} (r_{\lambda})  \right|, \\
    r_{\lambda} &= \left( \frac{8 \pi G A}{3 \lambda} \right)^{2/3}.
\end{align}
For a pure power-law initial profile, the final profile reaches zero density at the tidal radius $r_t = 0.24 r_\lambda$ \citep{stuecker_2022}. If we set up a pure power-law profile and evaluate the structure of the remnant using the \textsc{adiabatic-tides} model, we find that its annihilation luminosity is given by
\begin{align}
    J(r > r_{\rm{min}}) &= 4 \pi A \log \left( \frac{\SI{7.759e-3}{} r_\lambda}{r_{\rm{min}}} \right), \label{eqn:Jcusp_smoothtide}
\end{align}
so that in this case the effect of smooth tide on the annihilation luminosity  is equivalent to a sharp truncation of the initial profile at $0.78\%$ of $r_\lambda$. 

To model accurately the joint effect of stellar encounters and tidal stripping, the whole encounter and tidal history should, in principle, be considered. Here, we will simply assume that we can model the joint effect approximately by first considering the effect of all stellar encounters -- by truncating the prompt cusp as described in Section \ref{sec:simulations} -- and thereafter applying the pericentre tidal field $\lambda$. We choose this order because it is technically simpler for us, but we expect that consideration of the effects in reverse order would be just as valid and would give similar results.

\begin{figure}
    \centering
    \includegraphics[width=\columnwidth]{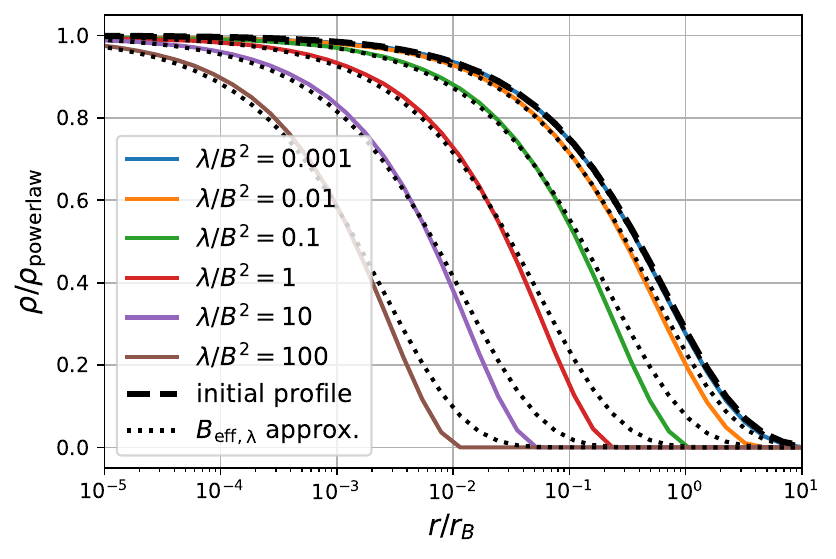}
    \includegraphics[width=\columnwidth]{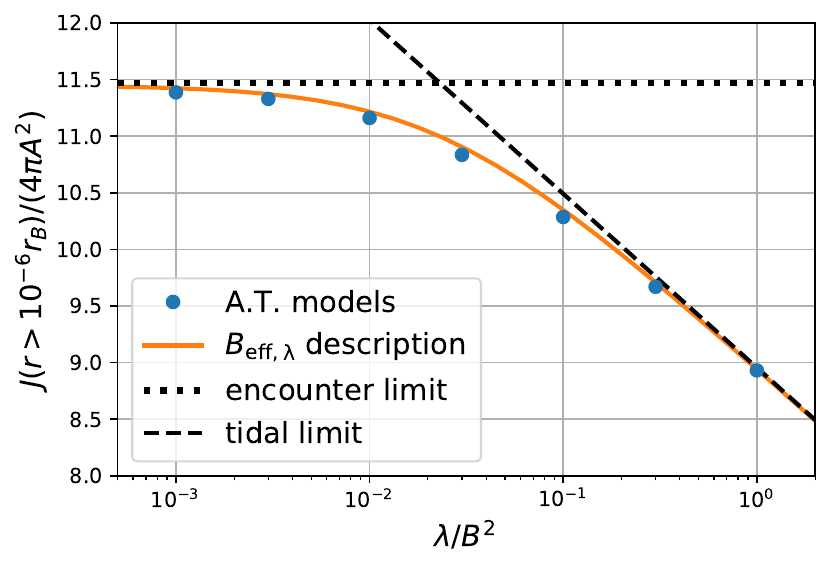}
    \caption{Top: Transfer functions for shocked power-law profiles that are subsequently adiabatically exposed to a tidal field with amplitude $\lambda$. The dashed line shows the initial (post-encounter) profile and the dotted lines show the effective description that we propose. Bottom: $J$-factors for the corresponding profiles. The  two limiting cases apply when either encounter truncation or  tidal truncation dominate. The intermediate regime  is described by an effective shock parameter $B_{\rm{eff}, \lambda}$.}
    \label{fig:tides_transfer}
\end{figure}

We set up a profile following the shocked power-law form,
\begin{align}
    \rho(r) &= A r^{-3/2} \exp(-1.256 (r/r_B)^{0.639}),
\end{align}
and we apply various tidal fields $\lambda$ using the \textsc{adiabatic-tides} model.  
We note that this set-up has one relevant parameter -- the ratio between the two truncation scales:
\begin{align}
    \frac{r_\lambda}{r_B} &= \left( \frac{B^2}{\lambda} \right)^{2/3}.
\end{align}
We note that the annihilation luminosity in equation \eqref{eqn:Jcusp_smoothtide} is equivalent to that in equation \eqref{eqn:jpowerlawapprox} for a single stellar encounter with shock parameter
\begin{align}
    B_{\rm{\lambda}} &= \sqrt{43.4 \lambda} \label{eqn:Blambda} \,\, .
\end{align} 
Thus, in the limit $B_\lambda \gg B$ (where the tidal field sets the truncation scale) we expect the annihilation luminosity to be  equivalent to that after a single shock with strength $B_\lambda$, while in the limit $B_\lambda \ll B$, it should be equivalent to a single shock of strength $B$. A reasonable guess for intermediate cases is for the joint effect to be given by a weighted average of the two scales,
\begin{align}
    B_{\rm{eff, \lambda}} &= \sqrt{B^2 + 42.2 \lambda}. \label{eqn:Befflambda_app}
\end{align}
In principle any $p$-norm of $B$ and $B_\lambda$ would be a reasonable guess:  we find that $p=2$ works very well.\footnote{$p=1.8$ actually works slightly better, but we stick to $p=2$ for simplicity.} In Figure \ref{fig:tides_transfer} we compare the transfer functions obtained from the \textsc{adiabatic-tides} calculations to an effective description assuming that the joint effect of encounters and tides is equivalent to a single shock with $B_{\rm{eff, \lambda}}$. This works reasonably well in the regime where $\rho/\rho_{\rm{powerlaw}} > 0.5$. In the regime where this ratio is smaller, the approximation is worse; the tidal truncation is complete at the tidal radius, whereas the encounter truncation has a much longer tail. However, the tail of the profile is almost irrelevant for the annihilation luminosity, and is unlikely to be correct in the \textsc{adiabatic-tides} description \citep{stuecker_2022}. 

The bottom panel of Figure \ref{fig:tides_transfer} shows the J-factors obtained by integrating $\rho^2$ for the adiabatic remnants over the range $[10^{-6} r_B, \infty]$. Note that we chose the lower bound so that it is well in the power-law regime of the profile for all the cases considered. Apart from that, it is, of course, arbitrary, and so the absolute offset on the $J$-axis is also arbitrary. We show additionally the two limiting cases of a pure encounter truncation and a pure tidal truncation, together with the effective description by a single shock with strength $B_{\rm{eff, \lambda}}$. Clearly this recovers the asymptotic limiting cases as well as the intermediate regime very well. 

\begin{figure}
    \centering
    \includegraphics[width=\columnwidth]{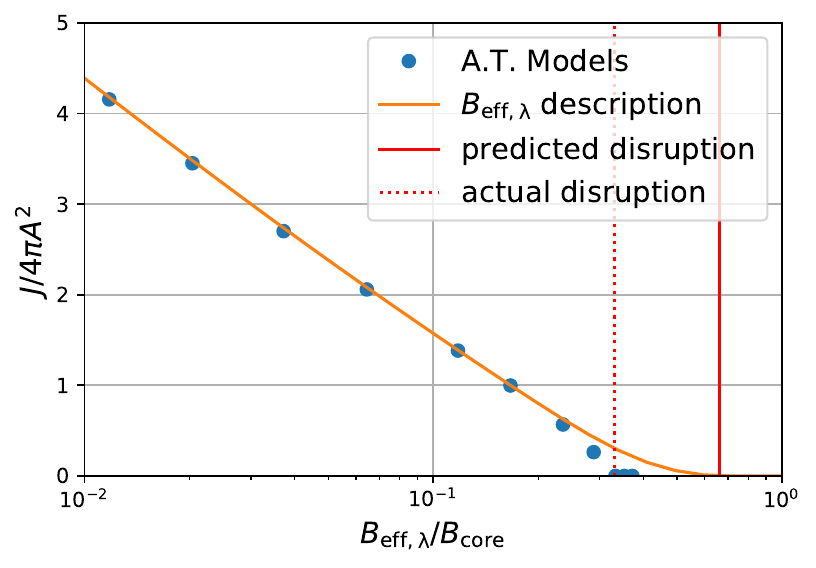}
    \caption{$J$-factors of cored prompt cusps that were adiabatically exposed to tidal fields of different amplitudes. The $J$-factors are very well approximated by the $B_{\rm{eff},\lambda}$-description, except for the regime $B_{\rm{eff}, \lambda} \gtrsim 0.2 B_{\rm{core}}$. However, such large tidal fields are anyways not found in the Milky Way.} 
    \label{fig:Jcore_smooth_tides}
\end{figure}

To verify that this gives a reasonable description of the joint effect of encounters and of the  smooth tidal field in all relevant cases, we must also check that it applies to cored initial profiles. Here, we only consider the case where $B_\lambda \gg B$ so that we do not have to deal with two scales at the same time, but only with a single scale $B_\lambda / B_{\rm{core}}$. We set up cored profiles as described in Section \ref{sec:phasespacecores} and apply tidal fields of varying amplitudes through the \textsc{adiabatic-tides} model. We show the corresponding $J$-factors in Figure \ref{fig:Jcore_smooth_tides}. 

Clearly the effective description works very well for cored profiles also, as long as $B_{\rm{eff}, \lambda} \lesssim 0.2 B_{\rm{core}}$. Beyond that scale the cored profiles reach an earlier disruption threshold of $B_{\rm{dis}, \lambda} = 0.33 B_{\rm{core}}$ which is a factor two smaller than the encounter disruption threshold. While it would certainly be possible to improve our effective model further to incorporate this reduced disruption threshold, this is unnecessary, since such large tidal fields are not reached in the Milky Way -- especially not at the large radii where smooth tides dominate over stellar encounters and  $B_\lambda \ll \SI{1}{\kilo \metre \per \second \per \parsec}$ typically. We conclude that treating the joint effect of stellar encounters and smooth tides as that due to a single encounter with a shock parameter $B_{\rm{eff}, \lambda}$ is an excellent approximation within the \textsc{adiabatic-tides} framework. We note that this may  slightly overestimate the effects of smooth tides, since the \textsc{adiabatic-tides} predictions are often a slight overprediction in practice \citep{stuecker_2022, aguirre_2022}. However, since the effects on cusp annihilation luminosity are in any case rather weak, this seems acceptable.


\bsp	
\label{lastpage}
\end{document}